# Objective Estimation of Spatially Variable Parameters of Epidemic Type Aftershock Sequence Model: Application to California

Shyam Nandan[1], Guy Ouillon[2], Stefan Wiemer[1] and Didier Sornette[3]

**Affiliations:**

[1]ETH Zürich, Swiss Seismological Service, Sonneggstrasse 5, 8092 Zürich, Switzerland

[2]Lithophyse, 4 rue de l'Ancien Sénat, 06300 Nice, France

[3]ETH Zürich, Department of Management, Technology and Economics, Scheuchzerstrasse 7, 8092 Zürich, Switzerland

**Corresponding Author:**

Shyam Nandan, ETH Zürich, Swiss Seismological Service, Sonneggstrasse 5, 8092 Zürich, Switzerland. (shyam4iiser@gmail.com)

**Key points:**

Efficient data driven method for estimation of spatially variable ETAS parameters.

Evidence for existence of triggering possibly through fluid-induced activation.

Evidence for seismic coupling independent of hypocentral depth.

**Abstract**

The ETAS model is widely employed to model the spatio-temporal distribution of earthquakes, generally using spatially invariant parameters. We propose an efficient method for the estimation of spatially varying parameters, using the Expectation Maximization (EM) algorithm and spatial Voronoi tessellation ensembles. We use the Bayesian Information Criterion (BIC) to rank inverted models given their likelihood and complexity, and select the best models to finally compute an ensemble model at any location. Using a synthetic catalog, we also check that the proposed method correctly inverts the known parameters.

We apply the proposed method to earthquakes included in the ANSS catalog that occurred within the time period 1981-2015 in a spatial polygon around California.

The results indicate significant spatial variation of the ETAS parameters. We find that the efficiency of earthquakes to trigger future ones (quantified by the branching ratio) positively correlates with surface heat flow. In contrast, the rate of earthquakes triggered by far-field tectonic loading or background seismicity rate shows no such correlation, suggesting the relevance of triggering possibly through fluid-induced activation. Furthermore, the branching ratio and background seismicity rate are found to be uncorrelated with hypocentral depths, indicating that the seismic coupling remains invariant of hypocentral depths in the study region.

Additionally, triggering seems to be mostly dominated by small earthquakes. Consequently, the static stress change studies should not only focus on the Coulomb stress changes caused by specific moderate to large earthquakes, but also account for the secondary static stress changes caused by smaller earthquakes.

**Introduction:**

The Epidemic Type Aftershock Sequence (ETAS) model [Kagan and Knopoff, 1981, 1987; Ogata, 1988, 1998] is a widely used statistical model to describe the occurrence of earthquakes in space, time and magnitude. In this model, any earthquake irrespective of its size can trigger other (larger or smaller) earthquakes, which in turn can trigger more earthquakes and so on, leading to a cascade of triggering. The key feature of the ETAS model is the apparent lack of traditional labels such as foreshock, mainshock and aftershock [Helmstetter and Sornette, 2003a; Helmstetter et al., 2003], which are often used for earthquakes by seismologists [see, for instance, Gardner and Knopoff, 1974; Reasenberg, 1985; Zaliapin et al., 2008], based on the parsimonious assumption that the same physical mechanisms give rise to all earthquakes.

Notwithstanding its simplicity, the ETAS model has been exceptionally successful in describing the numerous statistical properties associated with earthquakes [see for e.g. Helmstetter and Sornette, 2002a, 2002b, 2003a and 2003b; Helmstetter et al., 2003]. However, it fails to account for several key properties of seismicity such as existence of stress shadow regions (where seismicity rate following an earthquake is suppressed) [see for e.g. Nandan et al., 2016; Meier et al., 2014]; multifractal nature of spatial distribution of earthquakes [Kamer et al., 2013]; magnitude dependent exponent of Omori law [Ouillon and Sornette, 2005] and so on. Despite these failures, it has been very successful (relative to other models) in forecasting the rates of future events, to the extent that it easily outperforms the physics based models of seismicity and ranks among the best models of earthquake forecasting developed to date [Werner et al., 2011; Console et al., 2007; Iwata, 2010; Dieterich, 1994].

Considering that the parameters of the ETAS model are the manifestations of the physical properties of the crust, which exhibit spatial variability, investigating the possible existence of spatial variability of ETAS parameters is justified. In fact, numerous case studies [see Utsu and Ogata, 1995 for list of references; Wiemer and Katsumata, 1999] have documented the variability of several ETAS parameters. For instance, the exponent of the modified Omori law (an empirical law constituting the ETAS model) has been found to vary in a wide range (0.6-2.5) [Utsu and Ogata, 1995] and has been proposed to be related to the tectonic condition of the region such as structural heterogeneity, stress and temperature. In another application, Guo and Ogata [1997] reported the variation of the exponent of the aftershock productivity law (another empirical law contributing to the ETAS model) in the range 0.2-1.9 [Helmstetter, 2003].

Despite notable evidence of spatial variation in ETAS parameters, its model parameters are generally considered to be spatially homogenous [Zhuang et al., 2004; Werner et al., 2011; Helmstetter et al., 2006]. Such simplifications, mostly made for computational convenience, have overarching ramifications. For instance, based on the spatially invariant estimate of the aftershock productivity law exponent for the Southern California catalog, Helmstetter [2003] concluded that seismicity triggering is driven by small earthquakes. This result has profound implications as it casts doubt on most of the stress change studies, which consider static stress changes by only moderate to large earthquakes and ignore the Coulomb stress changes caused by smaller earthquakes to predict the location of future ones [King et al., 1994; Stein, 1999; Stein et al., 1994; Oppenheimer et al., 1988, Parsons and Dreger, 2000; Wyss and Wiemer, 2000; Bhloscaidh et al., 2014]. As a result, it is important to reinvestigate the findings of Helmstetter [2003] without the restrictive assumption that the ETAS parameters are spatially invariant in order to falsify (or confirm) her mean field observation and possibly differentiate regions where her result is valid from those where it is not.

Another consequence with potentially serious impacts of the assumption of spatially homogenous ETAS parameters is the inability to distinguish regions of smaller hazards from the higher ones. Considering the spatial variability of the parameters would not only clarify this issue but would further allow us to differentiate regions in terms of potential of the type of hazard (long term or short term), which could possibly aid in policy formulations.

There is also a general lack of understanding about the physical origin of ETAS parameters, which primarily arises from scarce attempts to correlate those parameters with geophysical measurements. Some of the noteworthy attempts include investigations of Kagan et al. [2010], Enescu et al. [2009] and Chu et al. [2011]. In particular, Kagan et al. [2010] and Chu et al. [2011] investigated the variation of ETAS parameters across different tectonic settings. On the other hand, Enescu et al. [2009] focused on the variation of the productivity law exponent and its correlation with surface heat flow measurements. A common aspect of these three works is the a priori delineation of the space-time window used to preselect the earthquakes that are modelled using a spatially homogenous ETAS model. For instance, Kagan et al. [2010] and Chu et al. [2011] use the definition of global tectonic zones proposed by Bird [2003] and Bird and Kagan [2004] to select earthquakes from the National Earthquake Information Center's (NEIC) Preliminary Determination of Epicenters (PDE) catalog to construct five sub-catalogs (depending on the location of earthquakes in one of the five types of tectonic zones). The authors then fit spatially homogenous ETAS model to each of the sub-catalogs of earthquakes to obtain ETAS parameters for each of the five tectonic zone types. It is important to note that the authors completely ignore the consequential possibility of spatial variability of the ETAS parameters within a given tectonic zone. This constitutes a simplification, which is hard to justify, not only because of the crustal heterogeneities but also because of the well known variability of far field tectonic loading at least at the scale of tectonic zones used in these studies. Moreover, the authors also make the unjustified assumption that the earthquakes occurring in one zone cannot trigger earthquakes in any other zone.

On the other hand, Enescu et al. [2009] use a window based method [Vidale and Shearer, 2006] to identify earthquake sequences that are "well" separated from other seismicity in space and time. Post identification, the authors fit a space independent spatially homogenous ETAS model to each of the seismic sequences individually. However, the window based pre-selection of earthquakes introduces a brutal cutoff in space and time beyond which earthquakes are considered to be independent in terms of triggering since inter-sequence earthquake triggering is assumed non-existent. As a result, all earthquake sequences are then thought to be independent of each other. On a large scale, for e.g. as studied by Kagan et al. [2010] and Chu et al. [2011], the finiteness of the catalog (in terms of space and time) might not be a major issue. However, at the smaller scales of individual earthquake sequences, as in the study of Enescu et al. [2009], serious biases could be introduced in the parameter estimates [Wang et al., 2010].

In the few scarce studies mentioned above, the adoption of such ad-hoc measures to quantify the spatial variability in the ETAS parameters stems primarily from the near absence of reliable methods to partition space and estimate model parameters. Recently,

Ogata et al. [2003] and Ogata [2004, 2011] has proposed a novel and systematic method to hierarchically estimate the spatial variation in all the parameters of the ETAS model. In his procedure, Ogata models an earthquake catalog with N earthquakes with nearly 5N parameters. On the surface, the proposed model seems to have a huge number of parameters. However, all 5N parameters are not all independent. Indeed, the effective number of parameters is determined by the strength of the roughness penalty imposed by the author. Even though, the model does not effectively have 5N parameters, one still has to invert for them. While the inversion problem would remain tractable for catalog with limited number of earthquakes, we foresee that the computation time (and amount of memory needed) would become extremely large as soon as one would want to apply this method to large earthquake catalogs. Moreover, it is not clear (due to lack of synthetic tests in Ogata [2004, 2011]) if this method reliably inverts the underlying ETAS parameters. We think that appropriate synthetic tests should be the minimum pre-requisite for any new method that attempts to invert parameters. Moreover, the burden to prove the reliability of the method lies on the shoulder of the proponent. While Ogata or others might have done synthetic tests, we were unable to find those in the public domain.

The goal of the present article is to present a method that can reliably invert the spatial variability of the ETAS parameters, thus providing novel information for improving our understanding of the physical origin of the parameters via the existence of correlations with geophysical properties of the crust.

The article is organized as follows. In section 2, we describe our new method to jointly invert the spatially variable parameters of the ETAS model. We also demonstrate, using appropriate synthetic tests, that our method is able to recover correctly the underlying spatial pattern of the parameters used to generate synthetic catalogs (see supplementary Text S1). As our data set, we use the catalog of earthquakes spanning the whole California, which is described in section 3. In section 4, we apply our method to this dataset and present our main results. In section 5, we investigate the origin of this spatial pattern by correlating the background and triggering seismicity parameters with some geophysical measurements such as surface heat flow and hypocentral depth. Finally, we present our conclusions and propose directions for future work in section 6.

## 2. METHOD:

### 2.1 Epidemic Type Aftershock Sequence (ETAS) Model:

As already mentioned, the ETAS model is actively used to model the spatio-temporal distribution of earthquakes [Zhuang et al., 2002; Helmstetter and Sornette, 2002; Daley and Vere-Jones, 2002]. The ETAS model is an adapted version for seismicity of the self-excited conditional Poisson process [Hawkes, 1971a, 1971b; Hawkes and Oakes, 1974]. In this model, the conditional seismicity rate, $\lambda(t, x, y | \mathcal{H}_t)$, at any location $(x, y)$ and time, $t$, depends on the history of the earthquake occurrences up to $t$ and is given by:

$$\lambda(t, x, y | \mathcal{H}_t) = \mu(x, y) + \sum_{i:t_i<t} g(t - t_i, x - x_i, y - y_i, m_i) \quad (1)$$

In Equation (1), $\mathcal{H}_t = \{(t_i, x_i, y_i, m_i): t_i < t\}$ represents the history of the process up to time $t$. $(t_i, x_i, y_i, m_i)$ respectively correspond to the time, *x*-coordinate, *y*-coordinate and magnitude of the $i^{th}$ earthquake in the catalog.

$\mu(x, y)$ is the background intensity function, which is assumed to be independent of time, while $g(t - t_i, x - x_i, y - y_i, m_i)$ is the triggering function. Several forms have been proposed for the triggering function [Console et al., 2003; Ogata, 1998; Zhuang et al., 2004; Zhuang et al., 2005; Harte, 2016]. In this paper, we use the form similar to the one proposed by Zhuang et al. [2005]:

$$g(x, y, t - t_i, x - x_i, y - y_i, m_i) = \frac{K e^{a(m_i - M_0)}}{\{t - t_i + c\}^{1+\omega} \{(x - x_i)^2 + (y - y_i)^2 + d e^{\gamma(m_i - M_0)}\}^{1+\rho}} \quad (2)$$

$\theta(x, y) = \{\mu, K, a, c, \omega, d, \gamma, \rho\}$ represents the set ETAS parameters, which can feature spatial variation. As already mentioned, in most previous works, $\theta(x, y)$ is generally assumed to be spatially invariant.

The triggering function is composed of several components:

1. The numerator $K e^{a(m_i - M_0)}$ represents the "fertility" or "productivity" of the "parent" earthquake. It is composed of a coefficient $K$ that may be space dependent and of an exponential function of the parent earthquake magnitude $m_i$. The exponent $a$ is the fertility exponent quantifying the relative productivity of earthquakes as a function of their magnitudes. Its value determines crucially the relative importance of small versus large earthquakes in their overall triggering impact (Helmstetter, 2002).

2. Spatial kernel, $\{(x - x_i)^2 + (y - y_i)^2 + d e^{\gamma(m_i - M_0)}\}^{-1-\rho}$, describes the spatial distribution of offspring around the $i^{th}$ earthquake. $d e^{\gamma(m_i - M_0)}$ measures the magnitude dependent spatial extent of the aftershock zone of the $i^{th}$ earthquake. Note that, while we have assumed for simplicity that the aftershock density around a mainshock decays with distance according to a simple power-law, there have been other studies [Gu et al., 2013; Moradpour et al., 2014], based on earthquake declustering method proposed by Zaliapin et al.

[2008], that suggest that the decay of the aftershock density with distance might not be a simple power-law.

3. Omori Kernel, $\{t - t_i + c\}^{-1-\omega}$, describes the temporal distribution of offspring following the $i^{th}$ earthquake, according to the modified Omori law [Utsu, 1995], as used to describe the rate of aftershocks.

4. 
$$G_i(\theta) = \int_{t_i}^{T} \iint_S g(t - t_i, x - x_i, y - y_i, m_i) dx\, dy\, dt \qquad (3)$$

gives the expected number of offspring of first generation with magnitude larger than a magnitude $M_0$ of triggering, of any earthquake with magnitude $m_i$ in the time period $[t_i, T]$ and in the spatial polygon S.

5. $M_0$ is the magnitude of the smallest earthquake that can trigger its own aftershocks [Sornette and Werner, 2005a]. For convenience, it is generally assumed that all earthquakes below the magnitude of completeness (see Sornette and Werner [2005b] for implications) of a catalog do not trigger any aftershocks [Ogata, 1988; Kagan, 1991; Ogata, 1998; Console et al., 2003; Ogata, 2004; Zhuang et al., 2004].

6. In order to express the Omori and Spatial kernel as probability density functions while computing the log-likelihood, the respective exponents $\omega$ and $\rho$ are constrained to be positive.

Conventionally, these parameters are obtained by maximizing the log likelihood given by:

$$l(\theta) = \sum_i \log\left(\lambda(t_i, x_i, y_i | \mathcal{H}_{t_i})\right) - \int_0^T \iint_S \lambda(t, x, y | \mathcal{H}_t)\, dx\, dy\, dt \qquad (4)$$

where [0,T] and S are respectively the time window and spatial polygon in which the data is observed.

## 2.2 Estimation of ETAS parameters using the Expectation Maximization (EM) approach:

As pointed out by Veen and Schoenberg [2008] and Schoenberg [2013], the maximum likelihood based inversion of ETAS parameters has several deficiencies. Typically, the Loglikelihood defined in Equation (4) is maximized using a numerical optimization routine, because no closed form solution is available. In cases where the log-likelihood function is extremely flat in the vicinity of its maximum, which could arise due to lack of sample information and/or parameter correlations [Harte, 2016], the numerical optimization routines have convergence problems and can be substantially influenced by arbitrary choices of the starting values [Veen and Schoenberg, 2008]. This problem can be further aggravated by the fact that the log-likelihood (equation 4) can be multimodal due to the underlying model or as a result of numerical inaccuracies.

Moreover, the maximum likelihood based inversion is extremely slow as it involves the estimation of $\int_0^T \iint_S \lambda(t,x,y|\mathcal{H}_t)\, dx\, dy\, dt$ for each guess of $\theta$ in the optimization routine. Since no analytical expression is available, the integration is performed numerically. However, numerical approximation of a spiky function in 3D is computationally expensive, and can lead to a sluggish estimation of $\theta$ [Schoenberg, 2013].

It is also important to note that there is a lot of missing information in a given recording of earthquake sequences. By construction, the ETAS model attributes probabilistic weights to each possible filiation of which previous earthquake triggered which following earthquakes, while the knowledge of this progeny structure is absent in any catalog. The degeneracy associated with the many possible filiation histories is the cause for the degeneracy of the likelihood function and the sloppiness of the estimated parameters [Brown and Sethna, 2003]. Given missing information, the Expectation Maximization method seems ideally suited to cope with it. Indeed, Veen and Schoenberg [2008] proposed using the Expectation Maximization scheme [Dempster et al., 1977; Baum et al., 1970; Hartley et al., 1958] for the estimation of $\theta$. The method of estimation can be broken down into two main steps:

1. Expectation step (or E-step): given the current guess of the parameters at $n^{th}$ step, $\hat{\theta}^n$, we first compute the probability that the $j^{th}$ earthquake is the offspring of the $i^{th}$ earthquake, $P_{i,j}^{(n)}$, using:

$$P_{i,j}^{(n)} = \frac{g(t_j - t_i, x_j - x_i, y_j - y_i, m_i | \hat{\theta}^n)}{\lambda(t_j, x_j, y_j | \mathcal{H}_{t_j}, \hat{\theta}^n)} \tag{5}$$

Using $P_{i,j}^n$, we can then estimate the total number of independent events, $\phi^{(n)}$, using:

$$\phi^{(n)} = \sum_{j=2}^{N} \left(1 - \sum_{i=1}^{j-1} P_{i,j}^{(n)}\right) + 1 \tag{6}$$

We can also estimate the total number of direct aftershocks triggered by the $i^{th}$ earthquake, $\psi_i^{(n)}$, using:

$$\psi_i^{(n)} = \sum_{j=i+1}^{N} P_{i,j}^{(n)} \tag{7}$$

2. Maximization step (or M-step): in this step, we maximize the complete data log-likelihood, $l_c^n(\theta)$, defined as:

$$l_c^n(\theta) = -\log\left(\Gamma(\phi^{(n)} + 1)\right) - \mu AT + \phi^{(n)} \log(\mu AT) +$$
$$\sum_{i=1}^{N} \left\{-\log\left(\Gamma(\psi_i^{(n)} + 1)\right) - G_i(\theta) + \psi_i^{(n)} \log(G_i(\theta))\right\} +$$
$$\sum_{j=2}^{N} \left\{\sum_{i=1}^{j-1} P_{i,j}^{(n)} \log\left\{\frac{g(t_j - t_i, x_j - x_i, y_j - y_i, m_i)}{G_i(\theta)}\right\}\right\} \tag{8}$$

where N, A and T are respectively the total number of earthquakes present in the catalog, the area of the spatial region over which the earthquakes in the catalog are distributed, and the total time span of the catalog. We refer the reader to Veen and Schoenberg [2008] for detailed explanation of the different terms composing the complete data log-likelihood, $l_c^n(\theta)$, defined in equation 8.

The new estimate of the ETAS parameters, $\hat{\theta}^{n+1}$, is obtained by maximizing $l_c^n(\theta)$ using a numerical optimization routine.

3. We repeat the steps 1 and 2 as long as $|l_c^{n+1}(\theta) - l_c^n(\theta)| > 10^{-4}$.

Veen and Schoenberg [2008] demonstrated with examples and synthetic tests that the EM algorithm is not only less susceptible to the poor initial guesses of the parameters compared to the conventional ML approach, but also yields superior estimates in the sense that the estimated parameters are less biased compared to the parameters estimated using the conventional ML approach. This is because the complete data log-likelihood defined in Equation 8 makes an optimal converging guess about the branching structure of the earthquake catalog using the triggering probabilities defined in Equation 5.

**2.3 Extension of the EM approach to estimate spatially variable ETAS parameters:**

We further extend the algorithm described in the previous section for estimating the spatially variable background seismicity rate, $\mu(x, y)$, and aftershock productivity parameters, $K(x, y)$ and $a(x, y)$. For the sake of simplicity, we have considered all other ETAS parameters, $\Theta = \{c, \omega, d, \gamma, \rho\}$, to be spatially invariant. We also assume that the spatial region containing the earthquakes consists of $q$ known subdomains, $S =$ $\{S_1, S_2, S_3, \ldots, S_q\}$, with respective areas, $A = \{A_1, A_2, A_3, \ldots, A_q\}$. Those subdomains are assumed to coincide with a Voronoi partition of the whole space. We further assume that $\mu(x,y), K(x,y)$ and $a(x,y)$ are piecewise constant functions over $S$: $\mu = \{\mu_1, \mu_2, \mu_3, \ldots, \mu_q\}$; $K = \{K_1, K_2, K_3, \ldots, K_q\}$ and $a = \{a_1, a_2, a_3, \ldots, a_q\}$. We assume for simplicity that the productivity of a source event depends solely on its magnitude and on the productivity parameters corresponding to the Voronoi cell in which it is located. Note that this assumption is reasonable if the size of each subdomain is significantly larger than the length of the largest event it contains, and if spatial variation are smooth at that scale.

The conditional seismicity rate at any location $(x, y)$ and time $t$ is now defined as:

$$\lambda(t, x, y | \mathcal{H}_t) = \mu_{f(x,y)} + \sum_{i: t_i < t} g_{f(i)}(t - t_i, x - x_i, y - y_i, m_i) \qquad (9)$$

In Equation (9), $\mu_{f(x,y)}$ is equal to the background rate in the spatial partition that contains that location $(x, y)$ and $g_{f(i)}(t - t_i, x - x_i, y - y_i, m_i)$ is the triggering kernel corresponding to the earthquake of magnitude, $m_i$, which occurs at location $(x_i, y_i)$ at time $t_i$ and is enclosed in the spatial partition $S_{f(i)}$, where $f(x, y)$ and $f(i)$ are the indexes of the spatial partition and can only attain values between 1 and q. The function $g_{f(i)}(t - t_i, x - x_i, y - y_i, m_i)$ is given by:

$$g_{f(i)}(t - t_i, x - x_i, y - y_i, m_i) = \frac{K_{f(i)} * \exp\{a_{f(i)} * (m_i - M_0)\}}{\{t - t_i + c\}^{1+\omega} * \{(x - x_i)^2 + (y - y_i)^2 + d * e^{\gamma * (m_i - M_0)}\}^{1+\rho}} \qquad (10)$$

In the above equation, $K_{f(i)}$ and $a_{f(i)}$ correspond to the productivity parameters in the spatial partition $S_{f(i)}$ in which the $i^{th}$ earthquake is located. Remember that the other ETAS parameters, $\Theta = \{c, \omega, d, \gamma, \rho\}$, are assumed to not vary in space.

To estimate $\Theta$, $\mu$, $K$ and $a$, we follow the EM scheme outlined in the previous section, using a new complete data log-likelihood defined as follows:

$$l_c^n(\Theta, \mu, K, a) = \sum_{m=1}^{q} \left\{ \begin{array}{c} -\log\left(\Gamma\left(\phi_m^{(n)} + 1\right)\right) - \mu_m A_m T + \\ \phi_m^{(n)} \log(\mu_m A_m T) \end{array} \right\}$$

$$+ \sum_{i=1}^{N} \left\{ \begin{array}{c} -\log\left(\Gamma\left(\psi_i^{(n)} + 1\right)\right) - G_i^{f(i)}(\Theta, K, a) + \\ \psi_i^{(n)} \log\left(G_i^{f(i)}(\Theta, K, a)\right) \end{array} \right\}$$

$$+ \sum_{j=2}^{N} \left\{ \sum_{i=1}^{j-1} P_{i,j}^{(n)} \log \left\{ \frac{g_i^{f(i)}(t_j - t_i, x_j - x_i, y_j - y_i, m_i)}{G_i^{f(i)}(\Theta, K, a)} \right\} \right\} \quad (11)$$

In Equation (11), $G_i^{f(i)}(\Theta, K, a)$ is the expected number of offspring with magnitude larger than $M_0$, of the earthquake of magnitude $m_i$ that occurred at location $(x_i, y_i)$, at time, $t_i$, and is enclosed in the spatial partition $S_{f(i)}$ (see Equation (3)).

However, in reality, the spatial partition $S$, over which $\mu$, $K$ and $a$ have been assumed to be piecewise constant, is unknown. We thus now outline the method, motivated by Kamer and Hiemer [2015], which we use to estimate the spatially variable $\mu$, $K$ and $a$ in the case of unknown spatial partitions:

1. We first assume that the total number of spatial cells in a given partition required to capture the variability of $\mu(x,y)$, $K(x,y)$ and $a(x,y)$ is $q$.

2. We thus divide the whole spatial region into $q$ Voronoi cells. To do this, we first draw q random points, which are distributed uniformly within the spatial polygon defined by Schorlemmer and Gerstenberger [2007]. We use these points to construct Voronoi partitions using the algorithm proposed by Barber et al. [1996]. As these partitions are constructed over an infinite 2D plane, we then compute an intersection of each of these q Voronoi partitions with the spatial polygon defined by Schorlemmer and Gerstenberger [2007] to obtain q Voronoi polygons.

3. We then estimate Θ, $\mu$, $K$ and $a$ using the EM algorithm outlined in the previous section (section 2.2) in conjunction with Equations (9-11).

4. We then repeat steps 2 and 3 several times ($N_{iter}$) with different realizations of the random distribution of the centers of the Voronoi cells. We store the estimates of $\widehat{\Theta}$, $\hat{\mu}$, $\widehat{K}$ and $\hat{a}$ along with the final value of the complete data log-likelihood $l_c^{final}(\widehat{\Theta}, \hat{\mu}, \widehat{K}, \hat{a})$ for each estimation.

5. We then compute the penalized log-likelihood for each of the $N_{iter}$ estimates using the following equation:

$$BIC(\widehat{\Theta}, \hat{\mu}, \widehat{K}, \hat{a}) = -2l_c^{final}(\widehat{\Theta}, \hat{\mu}, \widehat{K}, \hat{a}) + N_{par} \log(N) \qquad (12)$$

In the above equation, $N_{par}$ is the total number of free parameters, which is equal to *5q+5* (each Voronoi cell has 2 parameters for the Voronoi center + 1 parameter for the background seismicity rate + 2 parameters for the aftershock productivity; plus 5 other ETAS parameters independent of the cells). *N* is the total number of earthquakes in the catalog.

6. We repeat steps 2 to 5 with increasing values of *q* (from 1 to $N_v$=400), where $N_v$ is the maximum number of Voronoi cells that can be used to divide the region. The choice of $N_v$ depends on the judgment of the modeler.

The number *q* of Voronoi cells (complexity level) decides the complexity of the model that we use to fit the data. In order to choose the optimal complexity required to describe the data, we first compute the median BIC for each complexity level using the BIC's corresponding to the $N_{iter}$ estimates for a given complexity level. We refer to the number of Voronoi cells for which we obtain the minimum median BIC as the optimal complexity level. However, the models corresponding to the optimal complexity level might not be significantly better than models corresponding to other complexity levels in describing the data. To account for this, we define an optimal complexity range around the optimal complexity level by repeatedly testing the null hypothesis that the median BIC corresponding to the optimal complexity level is equal to the median BIC of other complexity levels against the alternative hypothesis that it is not. All the models corresponding to any complexity level for which the null hypothesis cannot be rejected are then considered along with the models of the optimal complexity level for further computation of an ensemble model. However, each of the selected model is weighted according to its BIC for the computation of ensemble model. The weight, $w_i$, corresponding to a given model $M_i$, is given by following equation:

$$w_i = \frac{e^{-\frac{BIC_i}{N}}}{\sum_{i=1}^{n_{tot}} e^{-\frac{BIC_i}{N}}} \qquad (13)$$

In equation (13), $BIC_i$ is the BIC corresponding to the $i^{th}$ selected model; $n_{tot}$ is the total number of selected models and $N$ is the total number of earthquakes present in the catalog.

The weighted averaging of the selected models, which lie in the optimal complexity range, ensures that our method is capable of finding continuous variations of the parameters in space (if they indeed show a continuous variation). On the other hand, if the spatial variation of the parameters does feature some discontinuities, our method could easily detect them as well. As a loose analogy, the Voronoi partitions are reminiscent of the Haar wavelet in the wavelet transform formalism: such a discontinuous wavelet can be used to decompose and reconstruct any given signal, may it be continuous or not.

**3. Data:**

We use the earthquakes ($M \geq 0, depth \leq 40\ km$) cataloged by the Advanced National Seismic System (ANSS) in the period from January 1, 1981 until July 5, 2015 enclosed in the RELM/CSEP collection polygon defined by Schorlemmer and Gerstenberger [2007] (Figure 1a) for the analysis.

Catalog incompleteness is one of the major problems in seismological studies. The origin of this incompleteness is generally attributed to the limited sensitivity and coverage of the Earth by station networks [Kagan, 2003]. The problem of completeness is generally addressed by considering a magnitude threshold ($M_c$) above which the frequency magnitude distribution follows the Gutenberg-Richter relationship [Woessner et al., 2005; Gutenberg et al., 1944]. In Figure 1(b), we show the global empirical frequency-magnitude distribution of the selected catalog. The solid black line in the figure shows the magnitude threshold ($M_c$=2.1) estimated using the method proposed by Clauset et al. [2009] for which we obtain $b_{val}$=0.95. This important statistical parameter quantifies the relative frequency of earthquakes with small vs large magnitudes and is used repeatedly in our analysis.

However, the catalog is not complete at the same level at all times and all locations. As a result, we need to estimate the joint spatio-temporal variation of $M_c$ in the chosen time period and spatial polygon. The full spatio-temporal analysis is beyond the scope of this paper. Instead, we make the conservative assumption that the catalog is complete above a magnitude threshold of 3 at all times and at all spatial locations. The two observations that justify this assumption are the following. First, an independent analysis by Werner et al. [2011] justifies this assumption for the RELM/CSEP collection polygon used in this study. Figure 2 of Werner et al. [2011] clearly shows that the magnitude threshold estimated using their method, which is a variation of the maximum curvature method proposed by Wiemer and Wyss [2000], is almost always smaller than our conservative assumption that $M_c \leq 3$. Note that Werner et al. [2011] used only the catalog until April 2010 for estimating the magnitude threshold. Further extending the catalog to 2015 should only further lower the estimates of the magnitude threshold due to the improving station coverage in the region. However, as the maximum curvature method underestimates the magnitude threshold [Mignan and Woessner, 2012] (see also Figure 1b), the magnitude threshold obtained by Werner et al. [2011] is likely an underestimation. Thus, in section 4.2, we further verify the abovementioned assumption that $M_c \leq 3$ is valid everywhere inside the study region using the more conservative estimator proposed by Clauset et al. [2009]. In doing so, we are also able to estimate the spatial variation of $b_{val}$.

Furthermore, we explore the time variation of $M_c$ estimated for the selected catalog within sliding time windows of size equal to one year using the method proposed by Clauset et al. [2009]. We clearly observe that $M_c$ seems to decrease with time (Figure 1c). This decrease of $M_c$ with time could be associated with the continuously improving sensitivity and coverage of the seismic network. We also observe that the estimated $M_c$ only seldom exceeds the conservative assumption of magnitude threshold ($M_c = 3$) made above, which further strengthens the validity of our choice $M_c \leq 3$.

Another important consideration while estimating the parameters of the ETAS model is the spatio-temporal boundary effect [Wang et al., 2010]. In our present paper, we have imposed a global spatio-temporal boundary, which is constituted by the collection of polygons proposed by Schorlemmer and Gerstenberger [2007] and the time period [1981-2015]. Any earthquake outside this spatio-temporal boundary is not allowed to contribute to triggering. This would certainly have an influence on the estimated parameters corresponding to earthquakes adjacent to the spatio-temporal boundary. However, this effect would only be limited because the parameters are estimated for the majority of earthquakes that are located away from the boundary. Note that such boundary effects could be easily accounted for by the use of auxiliary windows, as in Wang et al. [2010], and will be accounted for in future studies.

**4. Results:**

**4.1 Estimates of the ETAS parameters ($\mu$, $K$, $a$ and $\Theta$):**

We implement the algorithm proposed in Section 2.3 to estimate the spatially variable background seismicity rate and aftershock productivity parameters, as well as the other global ETAS parameters. We increase the number of Voronoi cells from $q = 1$ to $q = 480$. For each level of Voronoi complexity, we perform $N_{iter} = 200$ random partitions and store the solutions for all of them. We then rank all the solutions according to the penalized log-likelihood (BIC) score obtained using Equation (12). Figure 2a shows the BIC corresponding to all (96,000) solutions (black circles) as a function of the number of Voronoi cells used. The minima of the median BIC (shown as solid red line) corresponds to 286 Voronoi cells, which is indicated using a solid magenta line. The dashed magenta lines indicate the range of the number of Voronoi cells (214-384) for which inverted models are not significantly worse (or better) than the models corresponding to the optimal number of Voronoi cells. Note that we have computed this range by testing the null hypothesis that the median BIC for a Voronoi partition with $q$ cells is equal to the minimum median BIC against the alternative hypothesis that it is not, using the Wilcoxon Ranksum test at a significance level of 0.05. We further select all (34,200) solutions within the optimal range of Voronoi cells (214-384) to compute the ensemble model.

In Figures 2b-d, we show the weighted median estimates of $\mu$, $K$ and $a$ at the locations of the 21,448 earthquakes used to estimate these parameters. To obtain the weighted median estimate of $\mu$, $K$ and $a$ at the location of the earthquakes, we first assign to each earthquake the value of the estimated parameters $\mu$, $K$ and $a$ corresponding to the Voronoi cells within which the earthquake is located for each of the 34,200 selected solutions. Then, we use these solutions to compute the weighted median ensemble solution. The weight corresponding to each solution is computed using equation (13).

In Figure 3a-e, we show the variation of the estimates of each of the five parameters, $\Theta = \{c, \omega, d, \gamma, \rho\}$, as a function of number of Voronoi cells used. All the five parameters show systematic variation with increasing number of Voronoi cells. While the parameter $d$ decreases with increasing number of Voronoi cells, the other four parameters $c$, $\omega$, $\rho$ and $\gamma$ systematically increase with increasing number of Voronoi cells.

Using the the estimates of $\Theta$ corresponding to the individual selected solutions and the associated weights computed using equation (13), we compute the weighted median ensemble estimates of $\Theta$ and the complementary 95% confidence interval (shown using solid and dashed magenta lines respectively).

We find that the three parameters $\mu$, $K$ and $a$ show noticeable spatial variation (Figure 2b-d). To confirm that these spatial variations are indeed real and not artifacts of the inversion procedure, we quantitatively perform two synthetic tests in the Supplementary Text S1. First, we test if the inversion procedure introduces spurious spatial variation in the parameters even if they are spatially invariant (Text S1.1). Second, we test if the inversion procedure is unable to capture the correct patterns of spatial variability in the parameters (Text S1.2). We find that our method is able to "pass" both synthetic tests, which supports our claim that the observed spatial variation in the three parameters are real (see Supplementary Text S1).

### 4.2 Spatial variation of $M_c$ and $b$

We obtain the spatial variation of $M_c$ and $b$, shown in Figures 4a-c, using the Voronoi partitions corresponding to the selected 34,200 solutions. We first compute an individual $M_c$ and $b$ map for each of the 34,200 solutions and then obtain the weighted median ensemble estimates of $M_c$ and $b$ using the weights computed using the BIC corresponding to the selected solutions (see equation (13)). To compute an individual $M_c$ and $b$ map, we first group all earthquakes (M≥0) depending on which Voronoi cell they are enclosed in. We then estimate the $M_c$ and $b$ for each group of earthquakes. The b-value ($b$) for a group of earthquake is estimated using the following formula proposed by Tinti and Mulgaria [1987]:

$$b = \frac{\log\left(1 + \Delta M / (\overline{M} - M_c)\right)}{\log(10) * \Delta M} \quad (14)$$

In this equation, $\Delta M$ is the magnitude bin size that is used in the catalog to group the magnitude of the earthquakes; $\overline{M}$ is the average magnitude of the earthquakes with magnitudes larger than the assumed magnitude of completeness, $M_c$. Note that we set $\Delta M = 0.1$, following the general existing practice [Marzocchi and Sandri, 2003; Kamer and Hiemer, 2015].

$b$ relies heavily on the prior knowledge of $M_c$, which is unknown in general. We use two methods, the first one (known as Maximum Curvature method) was proposed by Wiemer and Wyss [2000] and the second one was proposed by Clauset et al. [2009], to estimate $M_c$. Note that originally, Clauset et al. [2009] proposed their method for power-laws. However, we would like to point to the reader that calibrating a power law p(x) is strictly identical to calibrating an exponential p(y) with the transformation y=ln(x). The Aki or Hill log-likelihood estimator is the same (by changing x with ln(x)). We refer the reader to Wheatley and Sornette [2015], where the authors make this point crystal clear for the application to extreme statistics.

From Figure 4a and 4b, we observe that $M_c$ estimated using the methods proposed by Wiemer and Wyss [2000], $\widehat{M}_c^2$, and Clauset et al. [2009], $\widehat{M}_c^1$, show noticeable variation in space. Both $\widehat{M}_c^2$ and $\widehat{M}_c^1$ seem to be systematically larger in offshore regions and Mexico. Larger incompleteness in these regions could be possibly attributed to poor station coverage. We also note that $\widehat{M}_c^1$ is systematically larger than $\widehat{M}_c^2$ (Figure 4d), which is consistent with the findings of other studies [Mignan and Woessner, 2012] that $\widehat{M}_c^2$ underestimates the magnitude of completeness. We find that the median value of the difference, $\widehat{M}_c^1 - \widehat{M}_c^2$, between the two estimates of $M_c$ is approximately 0.7 units. Using the maps of both $\widehat{M}_c^1$ and $\widehat{M}_c^2$, we are also able to justify our assumption that $M_c \leq 3$ is valid everywhere inside the study region with only few exceptions from the offshore region in Mendocino in the north and Mexico in the south.

We also find that *b* shows noticeable variations in space (Figure 4c). Some of these variations seem to be consistent with the ones reported in the literature [Tormann, 2011; Kamer and Hiemer, 2015; Tormann et al., 2014; Wiemer and Wyss, 2002]. For instance, as reported in these studies, regions such as the Mendocino Fault zone, the Cascadia mega thrust, the Parkfield section of San Andreas Fault, Northridge and so on are associated with low *b*. On the other hand, areas of high b-value on this map, like the region around Geysers, North Palm Springs, the creeping section of San Andreas fault, and so on have also been reported in these studies as regions with high *b*.

It has been often claimed that geothermal regions are exclusively associated with high b-values [see Wiemer and Wyss, 2002 and references therein]. It is interesting to note, however, that we find that, while some of the geothermal areas in the study region such as Geysers and North Palm Springs are associated with large b-values, several other geothermal areas such as the Coso geothermal field and Mammoth mountain are associated with moderate (~1) to low b-values (<0.9).

**4.3 Correlation among parameters**

In Figure 5a-f, we show the correlations between the parameters, which are estimated at the locations of the 21,448 M≥3 earthquakes. For clarity, we only plot the median value and 95% confidence interval (CI) of one set of parameter versus the median of the second set of parameter.

The general procedure to obtain one of these plots is the following. Given the spatially ensemble estimates of the parameters, each earthquake, $E_i$, can be associated with a pair of parameter values (say, $X_i$ and $Y_i$) depending on its location. We sort the earthquakes according to their corresponding $X_i$ values and divide the range of $X_i$ values into $k$ different bins, where $k$ varies between 1 and $n_{bin}$ (=50), where $n_{bin}$ is the total number of bins. Each of these bins is defined so that it contains the same number of earthquakes as each other bin. We consider the median of $X_i$ and $Y_i$ in the $k^{th}$ bin, as the representative parameter value of the $k^{th}$ bin. Then, we plot the median value and the 95% CI of $Y_i$ versus the median of $X_i$.

We find that, among all the parameter pairs, $K$ and $\alpha$ seem to have the strongest coupling, and are negatively correlated to each other. Such a strong coupling between $K$ and $\alpha$ is possibly due to the form of the aftershock productivity law prescribed in the ETAS model, in which both parameters can compensate each other in order to achieve a similar productivity. Indeed, a similar line of reasoning is presented in Harte [2016]. Harte [2016] argues that even though (a spatially invariant) ETAS model could have parameter space of 7 or 8 dimensions (depending on the spatial kernels assumed), the parameter values reside in a hyperplane of lower dimension. This can be seen from the simulations of Harte [2016, Table 2] where the eigenvalues of parameters obtained by refitting ETAS model to simulated data span ~2 less dimensions. This implies that correlations must exist between some of the parameters of the model. Indeed, this hypothesis is particularly verified by the negative correlation between α and κ in Harte [2016, Table 1]. The implication of the strong coupling of $K$ and $\alpha$ could be that these parameters cannot be correctly estimated without the prior knowledge of one of them. However, we demonstrate in the Supplementary Text S1.2 that, in spite of the strong coupling of the two parameters, our method is able to extract the correct spatial patterns of these two parameters from a synthetic dataset generated using the spatial patterns observed in the real data, without any prior knowledge of the any of the two parameters.

Another line of reasoning for the correlation between K and α could be based on Harte [2013]. It could be argued that, if an aftershock sequence(s) eventually dies out, then an ETAS model fitted to such data should have parameters in the stable regime. Models with parameters outside of this regime will be highly penalised by the log-likelihood (or other criteria) at the fitting stage. If the model is stable, then the expected number of all descendants of a given event must be finite. Following the notation of Harte [2013], it follows from equation 8 in Harte [2013] that $\kappa < 1 - \alpha/\beta$. Assuming that β is fixed, the preceding condition would imply that, if κ is large, then α must be small;

and if α is large and close to β, then κ must be small. Hence, they must be negatively correlated. However, we would like to point to the reader several assumptions (implicitly) made in the above mentioned chain of arguments that adds some doubts regarding the validity of above arguments. First, the condition $\kappa < 1 - \alpha/\beta$ in Harte [2013] derives from the assumptions that having a branching ratio smaller than 1 ($n < 1$) is a necessary requirement for the aftershock sequences to die out with probability one. However, while the aftershock sequences would necessarily die out when the $n < 1$, aftershock sequences have a finite probability to die out even when the n exceeds the critical value of 1. Second, the condition, $\kappa < 1 - \alpha/\beta$, also requires that there is no upper bound on the maximum magnitude, $M_{max}$, of earthquakes that can occur. Indeed, when there is a finite upper bound on $M_{max}$, the definition of n changes. In case of a truncated exponential distribution of magnitudes, the definition of n conditioned on whether $\alpha > \beta$, $\alpha = \beta$ or $\alpha < \beta$ is provided in the equations in Harte [2013, Appendix A]. As a result, the stability conditions are slightly relaxed. Third, β cannot be assumed to be fixed as it does feature spatial variation as can be seen from Figure 4c. Last but not least, the abovementioned arguments for a negative correlation between κ and α are further based on the equality assumption. Indeed, the negative correlation between κ and α would immediately follow if we knew a priori that $n = 1$ and is spatially invariant. In the subcritical regime ($n < 1$), κ can freely assume any value that does not violate the inequality constraint. The same is true if $n > 1$.

It is also important to consider that in our formulation of the ETAS model, we have assumed the c value of the Omori kernel to be independent of the magnitude of the mainshock while it might actually depend on it, either due to physical reasons [Dieterich, 1994; Narteau et al., 2002] or due to short term aftershock incompleteness [Hainzl, 2016; Helmstetter et al., 2006]. Such a simplification can possibly manifest itself in form of anti-correlation between $K$ and $\alpha$. In supplementary texts S2 and S3, we investigate this issue in detail. Using rigorous statistical test in supplementary text S2, we are able to show that the Omori kernel with a fixed c-value fits the observed decay rate of aftershocks in the real catalog very well. As a result, it cannot be rejected as a reasonable hypothesis. In supplementary text S3, we further modify our ETAS formulation such that the c-value of the Omori kernel depends on the magnitude of the mainshock, $c = c_0 e^{\eta(M_i - M_0)}$. Upon calibration of this modified ETAS model on the catalog used in this study, we find that the unmodified ETAS model (so far used in the study) describes the data as well as the modified ETAS model in terms of penalized log-likelihood (BIC) (see Figure S9). This implies that the gain in terms of BIC for the modified ETAS model over the unmodified one is non-existent. Furthermore, we also find that the spatially variable parameters $K, \alpha$ and $\mu$ obtained from the modified ETAS model are nearly equivalent to the ones obtained from the unmodified model. This automatically implies that the estimates of the parameters $K$ and $\alpha$ obtained from the modified ETAS model are also negatively correlated. Finally, we find that the value of parameter $\eta$ obtained with the modified ETAS model is -0.19. The negative value of $\eta$ indicates that the c value, which is thought to indicate the short term aftershock incompleteness duration, decreases with the magnitude of the mainshock. In fact, this observation is inconsistent with the hypothesis that short term incompleteness increases with the magnitude of the mainshock. While this later hypothesis might be true, it is not supported by the data when we consider only earthquakes with $M \geq 3$. The negative value of $\eta$ seems to be consistent with several physics based models such as the stress corrosion model [Scholz, 1968; Narteau et al., 2002] and rate and state friction model [Dieterich, 1994; Dieterich et al., 2000] hypothesizing that larger amplitudes of stress perturbations can lead to a decrease in the duration of the non-power-law regime in the rate of aftershock decay, which would imply that the c value of the Omori law would decrease with the magnitude of the mainshocks (it is assumed that larger earthquake would cause larger stress perturbations).

In the both discussions above, we tried to explain the negative correlation between $K$ and $\alpha$ as estimation artefacts. Yet, we failed to do so. It thus seems likely that the spatial patterns observed for $K$ and $\alpha$ in the real data are indeed genuine. It derives that the negative correlation is also real. We are not aware of any physical mechanism that can explain this negative correlation. It is possible that $K$ might be dependent on the local faulting density, while $\alpha$ might depend on the scaling of this density with the size of the aftershock zone [Helmstetter, 2003]. Thus, further understanding may come from local reconstructions of the fault network [Ouillon and Sornette, 2011; Wang et al., 2013; Nandan et al., 2016] coupled with physics-based models of stress transfer and rate-and-state friction [Dieterich, 1994].

**5 Discussion:**

**5.1 Branching Ratio:**

The branching ratio, $n$, defined as the average number of direct aftershocks per earthquake, is a key ETAS parameter. Based on the value of $n$, three ETAS regimes can be distinguished [Helmstetter and Sornette, 2002a]. The first regime corresponds to the case $n < 1$ and is also known as the subcritical regime. In this regime, aftershock sequences die out with a probability 1. The case $n > 1$ corresponds to the supercritical regime for which there is a non-zero probability that a given aftershock sequence grows exponentially without bounds. The case $n = 1$ corresponds to the critical regime, which separates the subcritical and supercritical regimes for which a rich set of critical behaviors of the triggered sequences can be expected [Saichev and Sornette, 2004; Saichev et al., 2005].

$n$ is given by the following equation:

$$n = \int_{M_0}^{M_{max}} G(m) * f(m) \, dm \tag{15}$$

In the above equation, $f(m)$ is equal to $\frac{b \, log(10)10^{-bm}}{10^{-bM_0} - 10^{-bM_{max}}}$, and describes the relative likelihood of an earthquake of magnitude $m$ to occur regardless of the location, time and magnitude of the parent-shock. $G(m)$ is the expected number of earthquakes triggered by an earthquake of magnitude $m$ and is computed using Equation (3). $M_{max}$ is the largest possible magnitude, while $M_0$ is the smallest magnitude of an earthquake that can trigger its own aftershocks [Sornette and Werner, 2005a].

In Equation (15), we still lack the spatially variable estimates of $M_{max}$ and $M_0$, which prevents us from estimating the spatially variable estimates of $n$. Nevertheless, we make the following simplifying assumptions to overcome this obstacle. First, we assume that both $M_{max}$ and $M_0$ are spatially invariant. Second, we assume that the largest possible magnitude that can occur in the study region is 8.5, based on previously reported values [e.g. Kagan, 1999]. Third, we assume that $M_0$ for the study region is equal to the minimum magnitude of the earthquakes present in the catalog used for the inversion of ETAS parameters, which in our case is equal to 3.

Figure 6a shows the obtained spatial distribution of $n$. We find that $n$ is far from uniform and varies within a wide range [0-1.2].

Prominent regions of high branching ratio (>0.8) include areas around Northridge, Hector mine and Landers earthquakes, the Parkfield section and the Santa Cruz

Mountain section of the San Andreas Fault Zone (SAFZ), Coalinga, Mammoth mountain, Coso geothermal fields, Geysers, Imperial Valley, Oceanside, Sugar Valley and so on. We also find that the Mendocino triple junction and creeping section of the SAFZ have locally anomalous branching ratios. While the Mendocino triple junction is characterized by higher branching ratios, the creeping section of the SAFZ is associated with smaller branching ratio, relative to its surrounding.

A relevant question is whether our estimated branching ratio is positively correlated with the local seismicity rate. In supplementary text S4, we answer this question in detail and show that our estimate of the branching ratio does not depend on the local seismicity rate.

Note that, in a few regions such as Mammoth mountain, $n$ locally exceeds the critical value 1, such that there would be a finite probability for the local seismicity to increase exponentially in the future. However, such exceedances necessarily have to be temporary, thus removing the physically improbable scenario of explosive seismicity. Seismicity sequences may indeed display apparent explosive behavior, which eventually subsides, following a scenario akin to "intermittent criticality" [Ben-Zion et al., 2003; Bowman and Sammis, 2004]. The hypothesis of temporary exceedance of the branching ratio is also consistent with the observations of Harte [2013, 2014] who provided evidence that productivity also varied temporally, between benign seismicity and highly active mainshock-aftershock sequences. This would imply that the ETAS parameters also vary in time, an aspect not considered in this study.

$n$ quantifies the efficiency of a given earthquake to trigger future earthquakes. Then, the existence of a significant spatial variability in $n$ indicates that the efficiency of earthquakes in triggering other earthquakes varies spatially. In the physical picture in which earthquakes trigger other earthquakes by pushing the almost critically stressed faults towards failure by adding miniscule stress perturbations on them, the variation in the efficiency of earthquake-earthquake triggering suggests that the crust is not equally critically stressed everywhere in the study region. This insight is supported by computational models encompassing both the long range and time organization of complex fractal fault patterns and the short time dynamics of earthquake sequences [Cowie et al., 1993; Sornette et al., 1994; 1995; Lee et al., 1999].

**5.2 Dominance of small or large earthquakes:**

The number of earthquakes of magnitude *m* scales as $10^{-b\,m}$ (Gutenberg-Richer law), and the number of earthquakes triggered by a typical earthquake of magnitude *m* scales as $10^{\alpha m}$ (fertility law, as formulated in ETAS model), where $\alpha = \frac{a}{log(10)}$. Therefore, the total number of earthquakes triggered collectively by all earthquakes of magnitude *m* scales as $10^{(\alpha-b)\,m}$. For $\alpha > b$, large earthquakes dominate triggering since $10^{(\alpha-b)\,m}$ is an increasing function of *m*: in this regime, a few very large earthquakes largely control the subsequent induced seismicity. For $\alpha < b$, small earthquakes dominate triggering since $10^{(\alpha-b)\,m}$ is a decreasing function of *m*: in this regime, the crowd of small earthquakes compensate for their relatively smaller individual triggering activity and, as a class, the small earthquakes dominate the overall seismicity triggering. For $\alpha = b$, all earthquake magnitude ranges contribute equally on average to the future triggered seismicity. Helmstetter [2003] reported empirical evidence that $\alpha < b$ for Southern California, suggesting that small earthquakes control seismicity triggering in this region. In the presence of spatially variable estimates of $\alpha$ and $b$, the picture becomes more complex as diverse regions can be found where large or small earthquakes dominate triggering.

Figure 6b shows the spatial variation of the weighted median estimate of $\alpha - b$. We find that the relation $\alpha \leq b$ holds for most parts of the study region. This indicates that triggering is either dominated by small earthquakes or small earthquakes play an equally dominant role as the large earthquakes in triggering in most of the study region. Our observations seem to be consistent with the results of not only Helmstetter [2003] but also with [Helmstetter et al., 2005; Marsan, 2005; Felzer et al; 2002, 2003; Gu et al., 2013] who also find, using the catalog of Southern California, that earthquake triggering is driven by small earthquakes. However, it should be noted that while these authors made the simplifying assumption that both $\alpha$ and $b$ were spatially invariant. In contrast, with spatially variable estimates of $\alpha$ and $b$, we are also able to identify localized regions where large earthquakes seem to dominate earthquake triggering. The most prominent among these localized regions with positive values of $\alpha - b$ are along the Mendocino fault zone and Cascadia megathrust.

Nevertheless, the general dominance of small earthquakes has strong implications for Coulomb stress change studies in the study region. Most of these studies, except for a few [e.g. Meier et al., 2014; Nandan et al., 2016], have focused on the Coulomb stress change caused by specific moderate to large earthquakes and completely ignored the secondary static stress changes caused by smaller magnitude aftershocks that seem to dominate the earthquake triggering, since $\alpha < b$. Taking account of secondary stress changes can possibly help explain why a significant fraction of aftershocks occur in stress shadow regions of the mainshocks (Felzer and Brodsky, 2005) and can thus help improve the forecasting skills of models based on Coulomb stress changes.

**5.3 Correlation of ETAS parameters with surface heat flow measurements**

Figure 7a-e shows the correlation between $n$, $\alpha$, $K$, $\mu$ and $b$ and local surface heat flow measurements.

To obtain these plots, we first smooth approximately 800 surface heat flow measurements in the study region (obtained from U.S. Geological Survey online heat flow database) to obtain heat flow estimates at the location of the 21,448 M≥3 earthquakes. Our smoothing method is the following. For a given spatial Voronoi partitioning scheme used during the fitting procedure, we first obtain the median heat flow estimate for each of the spatial cells using the enclosed heat flow measurements. All the earthquakes enclosed in each of the spatial cells are then assigned the corresponding median heat flow estimate. We repeat this two steps procedure for all the 34,200 Voronoi partition schemes corresponding to the selected solutions within the optimal complexity range (shown Figure 2a) to obtain 34,200 individual surface heat flow maps. Finally, we obtain the ensemble surface heat flow map by weighting all the individual surface heat flow maps using weights that are computed according to equation (13). We then choose the variable, say Y, whose correlation with surface heat flow we want to investigate. Each earthquake, $E_i$, is associated with a parameter value ($Y_i$) and a surface heat flow value ($HF_i$) depending on its location. We sort $Y_i$ according to the corresponding $HF_i$ value and divide the latter into $k$ different bins, where $k$ varies between 1 and $n_{bin}$(=50), where $n_{bin}$ is the total number of bins. Each of these bins is constructed so as to contain an equal number of earthquakes. We consider the median of $HF_i$ and $Y_i$ in the $k^{th}$ bin, as the representative variable values of the $k^{th}$ bin. Then, we plot the median value and the 95% CI of $Y_i$ versus the median of $HF_i$.

We find that the three parameters, $n$, $\alpha$ and $K$, show a systematic correlation with surface heat flow (Figure 7a-c). Both $n$ and $K$ first systematically increase with increase in surface heat flow (< 80 mW/m^2) and then saturate for higher heat flow values. On the other hand, $\alpha$ decreases with increasing heat flow values. We also find that the remaining two parameters, $\mu$ and $b_{val}$, show no systematic correlation with surface heat flow.

Our observation of a negative correlation between $\alpha$ and surface heat flow is consistent with the results reported by Enescu et al. [2009], who inverted the value of alpha for many earthquake sequences. Enescu et al. [2009] argued that a negative correlation of $\alpha$ and surface heat flow is consistent with the damage rheology model of Ben-Zion and Lyakhovsky [2006], which predicts that aftershock productivity is proportional to the effective viscosity in a region. According to Enescu et al. [2009], a decrease in the value of the productivity exponent is interpreted as a decrease in the aftershock productivity of earthquakes, and this decrease is expected with an increase in surface heat flow, which lowers the effective viscosity of the crust. However, the argument of Enescu et al. [2009] incorrectly identifies the decrease of the productivity exponent $\alpha$ with that of the productivity itself. Indeed, according to the ETAS model, the productivity $K \cdot 10^{\alpha m}$ of an earthquake of magnitude $m$ is also influenced by the pre-factor K and not just $\alpha$. We find that the large variations of $K$ influence the values of the branching ratio $n$ more than do the relatively small variations of $\alpha$. In other words, the prefactor $K$ plays a dominant role in dictating the aftershock productivity. Contrary to the findings of Enescu et al. [2009] and Yang and Ben-Zion [2009] and the predictions of the damage rheology model of Ben-Zion and Lyakhovsky [2006], we find that regions with high heat flow are more productive than regions with low heat flow in terms of aftershock generation, which is indicated by a systematic increase of $n$ and $K$ with increase in surface heat flow. The decrease of $\alpha$ is too small to have a countervailing effect.

The systematic increase of the branching ratio with increase in surface heat flow and its convergence to the critical value of 1 indicates that earthquake triggering is increasingly efficient in regions of high surface heat flow. Such an increase could indicate that the crust is closer to the local critical stress threshold associated with triggering in the regions of higher heat flow. This interpretation is supported by overwhelming evidence of remote dynamic triggering in areas with geothermal/volcanic activity in numerous case studies from around the globe (see table 2 of Hill and Prejean, 2007 for the list of reported cases). In fact, many of the areas with volcanic/geothermal activity, such as Geysers, Coso, Long Valley, Mammoth Mountain and Salton Sea area, with reported evidence of remote dynamic triggering, are part of our study region and are unambiguously associated with high values of branching ratio ($n > 0.9$). Assuming that the crust is stressed close to criticality in these geothermal/volcanic areas of very high heat flow allows us to reconcile both observations of remote dynamic triggering and high values of branching ratio. As the crust is very close to the local critical stress threshold necessary for triggering, even a small nudge provided by a miniscule stress change (static/dynamic) from a far-field source may be able to push some of the existing faults towards failure. In the same way, stress perturbations (static/dynamic) caused by an earthquake in its vicinity could efficiently trigger more earthquakes by nudging the surrounding faults, which are already close to unstable, towards failure.

However, while it is increasingly efficient for earthquakes to trigger other earthquakes (indicated by positive correlation between $n$ and surface heat flow), triggering by the far-field tectonic loading remains uncorrelated with surface heat flow (indicated by no correlation between $\mu$ and surface heat flow in Figure 7d). These two observations in combination point towards a dynamic weakening process rather than a static weakening of the crust that preferably occurs in the region of high surface heat flow. Considering that areas of very high surface heat flow in the study region, such as Geysers, Coso,

Long Valley, Mammoth Mountain, Salton Sea area and so on, are also very rich in fluids, we propose, in accordance with several researchers [Brodsky et al., 1998; Moran et al., 2004; Hill et al., 1995; Hill et al., 2002; Manga and Brodsky, 2006], that triggering in those regions could indeed be driven through dynamic excitation of crustal fluids. Passage of seismic waves from the distant large earthquakes or from those triggered locally by the far field tectonic loading could redistribute the pore pressures by changing the crustal permeability, for instance by disrupting clogged fractures and via hydraulic fracturing. As proposed by Hill and Prejean [2007], the pore-pressure redistribution mechanism may be particularly relevant in active geothermal areas, such as the Geysers and Coso geothermal fields, as fractures are sealed and high-pressure compartments form over relatively short timescales as minerals are precipitated from hot brines. This process of pore pressure redistribution may modify the Coulomb failure function such that the effective normal stress is decreased sufficiently to trigger failure [Cocco and Rice, 2002] or that quasi-static (aseismic) strains associated with local, fluid-driven deformation are sufficient to trigger earthquakes. Several other mechanisms, involving bubble excitations [Manga and Brodsky, 2006], magmatic intrusions or sinking crystal plumes [Manga and Brodsky, 2006], which have been proposed to account for the readily available evidence of dynamic triggering in geothermal areas could also be equally relevant in explaining the observation of highly efficient earthquake-earthquake triggering in the areas of very high surface heat flow in the study region.

Despite the overwhelming evidence of correlation between heat flow and branching ratio, it is important to note that we have completely ignored the effect of anthropogenic activities, such as fluid injection (and extraction) in the geothermal regions, on the seismicity in our analysis. Recently, Trugman et al. [2016] and Brodsky and Lajoie [2013] have shown convincing evidence that the background seismicity rate, $\mu$, shows systematic correlation with fluid injection and fluid extraction rates and varies in time. Ignoring the time variation of $\mu$ (and possibly of other parameters) can lead to systematic bias in the estimates of the parameters of the ETAS model. For instance, a geothermal region can display a constantly increasing seismicity rate with time due to constantly increasing fluid injection rates. An ETAS model with time invariant parameters could possibly characterize such a seismicity sequence as explosive, in which case the estimated branching ratio would be erroneously estimated as larger than 1. Furthermore, ETAS parameters (especially $\mu$) in regions with seismic swarms could also feature temporal variations, which when ignored could lead to biases in the estimated ETAS parameters [Jacobs et al., 2013; Hainzl et al., 2013; Kumazawa and Ogata, 2014].

**5.4 Correlation of ETAS parameters with Depth:**

Depth at which earthquakes occur is often suggested to be a controlling factor for their occurrence. In particular, based on arguments borrowed from the rate and state friction model [Dieterich, 1994], Scholz [1998] proposed a synoptic model of the variation of the frictional stability parameter, $\zeta = a_f - b_f$ (where $a_f$ and $b_f$ are here the parameters quantifying material properties in the Rate and State dependent friction law), as a function of depth for crustal faults and subduction zone interfaces. Scholz [1998] proposed that $\zeta$ is positive (indicating stable slip regimes) at shallow depths because of the presence of unconsolidated granular material, and at large depths because of the onset of plasticity at, and above, a critical temperature. Between the two stable slip regimes exists the unstable regime, for which $\zeta$ exceeds a certain threshold (see equation (2) in [Scholz, 1998]). This unstable regime corresponds to the seismogenic depth range over which earthquakes may nucleate. Motivated by these propositions, we investigate if the parameters of the ETAS model, in particular $\mu$ and $n$, are correlated with hypocenter depths.

To investigate these correlations, we first extend our method to invert spatially variable parameters in 3D. The whole procedure of the inversion (see section 2.3) remains the same, except for the following changes. First, we perform the spatial partitioning in 3D using Voronoi volumes. Second, we modify the spatial component of the triggering kernel such that it depends on hypocentral distances rather than epicentral distances.

It is interesting to note that the number of Voronoi cells for which we achieve the minimum median BIC is nearly the same for 2D (286 cells) and 3D (320 cells) inversions respectively. This suggests that very few new spatial cells are needed to explain the variation of the parameters along the newly added depth dimension, which possibly indicates that variation of the inverted parameters along depth is much smaller (or possibly non-existent) compared to their lateral variation.

We further try to systematically quantify the variation of $\mu$ and $n$ as a function of hypocentral depth in the following manner. First, we subtract the 2D estimates of $\mu$ and $n$ from their corresponding 3D estimates, both of which have been obtained at the location of all 21,448 M≥3 earthquakes used for the inversion of these parameters, in order to obtain respective parameter residuals, $R_\mu$ and $R_n$, at each location. In doing so, we remove the effect of the lateral variation of the parameters from the 3D estimates of $\mu$ and $n$, leaving behind $R_\mu$ and $R_n$. Any systematic variation (with depth) of the two residuals can then be purely attributed to the hypocentral depths.

In Figure 8a-b, we show the correlation of $R_\mu$ and $R_n$ with hypocentral depths. For clarity, we only show the median value and 95 % CI of $R_\mu$ and $R_n$ in different depth bins. We follow the same binning procedure as defined in the previous section 5.3. Both $R_\mu$ and $R_n$ show no systematic trend with hypocentral depth, indicating that $\mu$ and $n$ are nearly independent of hypocentral depth.

Our observations are in direct contradiction with the synoptic model proposed by Scholz [1998], which predicts a mid-crustal unstable regime (with efficient earthquake nucleation, propagation and triggering) sandwiched between upper and lower stable regimes. Our results rather suggest that both the far-field tectonic loading and stress perturbations caused by earthquakes could be equally efficient in nucleating unstable slip at all depths at a given horizontal location. While Scholz [1998] suggests that seismic coupling is strong in the middle part, and weak at shallow and large depths, our results indicate that seismic coupling seems constant along depth, even if it varies laterally in the study region. This thus suggests that the normal stress has no real effect on earthquake nucleation and that temperature also has no effect. This reinforces the conclusion that high heat flow areas show a singular behavior because of their fluid content, not because of higher temperatures.

It is interesting to note that, without the residual analysis, one does observe a global depth variation of $\mu$ and $n$, shown in Figure 8c-d. For clarity, we only show the median value and 95 % CI of $\mu$ and $n$ in different depth bins. We find that, globally, both $\mu$ and $n$ show a slight increasing tendency up to ~10 km depth, and then consistently decrease afterwards. These observations seem to be consistent with the model proposed by Scholz [1998]. However, in combination with the residual analysis, our observations indicate that the global depth variation of $\mu$ and $n$ observed for the whole study region is merely a geometrical effect, which only originates when we stack local $\mu$ and $n$ values observed in all the sub-regions (composing the whole study region) together, despite the fact that both $\mu$ and $n$ show no correlation with depth locally. The origin of the global depth variation of $\mu$ and $n$ as a geometrical effect rather than a physical effect can be explained if we consider that the maximum seismogenic depth varies laterally due to variation in surface heat flow.

Last but not least, we have not considered the influence of uncertainties in the location and magnitude of the earthquakes on the estimates of the ETAS parameters in this paper. Yet, we can be concerned by the possibility that such location uncertainties (especially along depth, which are usually larger) might blur any variation with location, and even totally overprint it if the uncertainties are large enough. Thus, the question of a possible influence of location (depth in particular) and magnitude uncertainties on the estimates of ETAS parameters is still open and should be considered in future studies.

## 6. Summary and Conclusions

We proposed a data-driven method to estimate the spatially variable parameters of the ETAS model. Our method is an improved extension of the stochastic declustering method proposed by Veen and Schoenberg [2008], which allows us to obtain the optimal spatially varying background seismicity rate as well as spatially varying estimates of other parameters of the ETAS model. The success of our method is demonstrated by the correct inversion of the parameters of a "realistic" synthetic catalog.

Applying our methodology to the earthquakes ($M \geq 3, depth \leq 40\ km$) cataloged by the Advanced National Seismic System (ANSS) in the period from January 1, 1981 until July 5, 2015 enclosed in the RELM/CSEP collection polygon, we obtained the spatial variability in the background seismicity rate $\mu$, the two productivity parameters ($K, \alpha$) and spatially invariant estimates of the remaining ETAS parameters. In addition, we obtained the spatial variability of the exponent of the Gutenberg Richter law ($b$) and magnitude of completeness ($M_c$). Using the spatially variable estimates of $K, \alpha$ and $b$, we obtained two derived ETAS parameters, the branching ratio ($n$) and $\alpha - b$, which respectively quantify the efficiency of earthquake-earthquake triggering and the dominance of large earthquakes relative to small earthquakes in their triggering contributions.

Based on the spatial variation of $n$, we deduced that the efficiency of earthquake-earthquake triggering is far from uniform in the study region, possibly due to the crust not being equally critically stressed everywhere. On the other hand, spatial variation of $\alpha - b$ indicate that triggering is mostly dominated by smaller earthquakes in the study region, with small pockets of equal or larger dominance of larger earthquakes in triggering. The widespread dominance of the smaller earthquakes in triggering (in the study region) necessitates the use of often ignored secondary stress changes (Coulomb stress changes caused by smaller earthquakes) in addition to the stress changes caused by larger earthquakes in static stress change studies.

Investigations of correlation of the branching ratio and the background seismicity rate with surface heat flow suggests the existence of triggering possibly through fluid-induced activation. The evidence of fluid mediated triggering is further accentuated with evidence of aftershock diffusion in areas with high fluid content.

Last but not least, we find that triggering and nucleation of earthquakes show no true correlation with hypocentral depths. The global correlation of background seismicity rate and branching ratio with hypocentral depths is rather a geometrical effect arising from the superposition of locally uniform depth dependences in a crust with a laterally varying seismogenic depth.

Our present work opens the following areas for further research.

First, we find that the number of background earthquakes, among the 21,448 (M≥3) earthquakes used for inversion, systematically increases with the number of Voronoi cells used to partition the area under study (Figure 9). For instance, the median number of background earthquakes increases from ~2100, for the minimum complexity (whole region treated as one cell), to ~4500, for 480 cells Voronoi cells. The median number of background earthquakes corresponding to the ensemble model is equal to ~4200. Relative to the ensemble model, the minimum complexity model (with spatially invariant parameters) underestimates the total number of background earthquakes by a factor of ~2 (relative bias= -50%). So, the ensemble model not only captures the optimal spatial variability of background seismicity rate but also its net amplitude. We speculate that both these factors should lead the ensemble model of background seismicity rates to outperform the long term forecast of spatially homogenous (or arbitrarily complex) ETAS models. We propose to test this hypothesis in our future work. Moreover, our method also allows us to distinguish regions where the triggering of earthquakes is extremely efficient from those where it is low (high $n$ vs. low $n$). This distinction can potentially improve short term forecasting of aftershocks, which constitute nearly 80% of the total observed seismicity (M≥3) in the study region, relative to models with a spatially homogenous branching ratio. This will also be tested in future work.

Second, as the ETAS parameters are correlated with each other in the calibration process, the assumption of spatial homogeneity for some of them might introduce biases in the estimates of spatially variable parameters. As a result, we need to extend our method to jointly invert the spatial variation of all the ETAS parameters.

Third, the ETAS parameters at a given location might not be stationary in time. This is especially relevant for regions with swarm activities and regions with anthropogenic activities (such as, fluid injection and extraction) leading to earthquake triggering. So, we propose to extend our method to jointly invert the spatio-temporal variation in the parameters, with special focus on geothermal regions.

Finally, in our current method, we have assumed that the ETAS parameters are source dependent (i.e. depend on the location of the source) for computational simplicity. Even though this assumption is reasonable if the size of each subdomain is larger than the length of the largest event it contains, and if spatial variations are smooth at that scale, a more physical description of the spatial variability of the ETAS parameters would be to assume that the parameters are target dependent. In the future, we will also explore this avenue and possibly compare the estimates of the parameters obtained from a target based approach to the present source based approach.

**Acknowledgment**

S.N. wishes to thank Y. Kamer, J.D. Zechar and J. Woessner for participating in discussions during the development of the work; U. Mannu for invaluable suggestions for improving the manuscript and S. Wheatley for suggesting new directions for future works. All the authors are also thankful to Yehuda Ben-Zion, David Harte, the associate editor and an anonymous reviewer for many invaluable suggestions, which helped to improve the manuscript.**References:**

**Table 1:** Description of frequently used symbols in the manuscript.

| Symbol | Description |
|---|---|
| $\lambda(t,x,y\|\mathcal{H}_t)$ | Seismicity rate at location $(x,y)$ and time, $t$, conditioned upon the history ($\mathcal{H}_t$) of the earthquake occurrences up to $t$. |
| $\mathcal{H}_t = \{(t_i, x_i, y_i, m_i): t_i < t\}$ | History of the earthquake occurrences up to time $t$; $(t_i, x_i, y_i, m_i)$ respectively correspond to the time, $x$-coordinate, $y$-coordinate and magnitude of the $i^{th}$ earthquake in the catalog. |
| $\mu(x,y)$ | The background intensity function, which is assumed to be independent of time. |
| $g(t-t_i, x-x_i, y-y_i, m_i)$ | The triggering function. |
| $Ke^{a(m_i-M_0)}$ | The "fertility" or "productivity" of the "parent" earthquake, with magnitude $m_i$ above the magnitude threshold $M_0$. |
| $M_0$ | Magnitude of the smallest earthquake that can trigger its own aftershocks. For convenience, it is set equal to the magnitude of completeness. |
| $\{t-t_i+c\}^{-1-\omega}$ | Omori Kernel, which describes the temporal distribution of offsprings following the $i^{th}$ earthquake. |
| $\{(x-x_i)^2 + (y-y_i)^2 + de^{\gamma(m_i-M_0)}\}^{-1-\rho}$ | Spatial kernel, which describes the spatial distribution of offsprings around the $i^{th}$ earthquake. |
| $G_i(\theta)$ | The expected number of offsprings of first generation with magnitude larger than a magnitude $M_0$ triggered by an earthquake with magnitude $m_i$ in the time period $[t_i, T]$ and in the spatial polygon S. |
| $\theta = \{\mu, K, a, c, \omega, d, \gamma, \rho\}$ | The set of spatially variable ETAS parameters. |
| $l(\theta)$ | Conventional log likelihood. |
| $P_{i,j}^{(n)}$ | The probability that the $j^{th}$ earthquake is the offspring of the $i^{th}$ earthquake, obtained in the $n^{th}$ Expectation (E) step. |
| $\phi^{(n)}$ | The total number of independent events obtained in the $n^{th}$ E step. |
| $\psi_i^{(n)}$ | The total number of direct aftershocks triggered by the $i^{th}$ earthquake, obtained in the $n^{th}$ E step. |
| $l_c^n(\theta)$ | complete data log-likelihood obtained in the $n^{th}$ Maximization (M) step. |
| $\hat{\theta}^{n+1}$ | The new estimate of the ETAS parameters obtained by maximizing $l_c^n(\theta)$ using a numerical optimization routine in the $n^{th}$ M step. |
| $\Theta = \{c, \omega, d, \gamma, \rho\}$ | ETAS parameters assumed to be spatially invariant in this study. |
| $S = \{S_1, S_2, S_3, \ldots, S_q\}$ | q spatial partitions in which $\mu, K$ and $a$ are piecewise constant functions. |
| $\mu_{f(x,y)}$ | The background rate in the spatial partition that contains that location $(x,y)$ |
| $g_{f(i)}(t-t_i, x-x_i, y-y_i, m_i)$ | The triggering kernel corresponding to the earthquake of magnitude $m_i$, which occurs at location $(x_i, y_i)$ at time $t_i$ and is enclosed in the spatial partition $S_{f(i)}$ |
| $f(x,y)$ and $f(i)$ | Indexes of the spatial partition that contains the locations (x,y) and $(x_i, y_i)$ respectively. Can only attain values between 1 and q. |
| $K_{f(i)}$ and $a_{f(i)}$ | The productivity parameters in the spatial partition $S_{f(i)}$ in which the $i^{th}$ earthquake is located. |
| $l_c^{final}(\hat{\Theta}, \hat{\mu}, \hat{K}, \hat{a})$ | The final value of the complete data log-likelihood. |
| $BIC(\hat{\Theta}, \hat{\mu}, \hat{K}, \hat{a})$ | The penalized log-likelihood. |
| $w_i$ | Weight of the $i^{th}$ selected model. |
| $b; M_c$ | Exponent of the Gutenberg-Richter law; Magnitude of completeness of the catalog. |
| $\widehat{M}_c^1; \widehat{M}_c^2$ | Magnitude of completeness estimated using the method proposed by Clauset et al. [2010] and maximum curvature method of Wiemer and Wyss [2000] respectively. |
| $n$ | The branching ratio, defined as the average number of direct aftershocks per earthquake. |
| $R_\mu$ and $R_n$ | Difference between the 2D estimates of $\mu$ and $n$ from their corresponding 3D estimates obtained at the location of all the earthquakes in the catalog. |

# Figure Captions

**Figure 1**: **(a)** Spatial distribution of earthquakes with magnitude larger than 0 that occurred within the time period 1 January 1981 to 5 July 2015 in the RELM polygon defined by Schorlemmer and Gerstenberger [2007] **(b)** Frequency magnitude distribution of earthquakes shown in the left panel; green circles show the number of earthquakes with magnitude larger than M; magenta stars show the number of earthquakes in magnitude bins of size 0.15 unit; the continuous black line shows the overall magnitude of completeness ($M_c$=2.1) estimated using the method proposed by Clauset et al. [2009]; the dashed red line shows the overall magnitude of completeness ($M_c$=1.1) estimated using the maximum curvature method [Wiemer and Wyss, 2000] **(c)** Time Series of $M_c$ estimated from the earthquakes shown in the left panel within sliding time windows of size 1 year using the method proposed by Clauset et al. [2009]; the horizontal continuous black line shows the magnitude threshold of 3, which is assumed to be the magnitude of completeness for this study; dashed black line shows the decreasing trend in the time series of $M_c$.

**Figure 2:** **(a)** BIC corresponding to 96,000 solutions as a function of the number of Voronoi cells used is shown using black circles; the median BIC corresponding to each Voronoi complexity level is shown using a solid red line; the continuous magenta vertical line corresponds to the minima in the median BIC curve and indicates the optimal complexity level; the dashed magenta lines indicate the optimal complexity range in which the median BIC for a given complexity level is not significantly different from the minimum median BIC. **(b-d)** Spatial variation of the **(b)** background seismicity rate ($\mu$, # earthquakes/$km^2/day$) **(c)** pre-factor of the aftershock

productivity (K) **(d)** exponent of the aftershock productivity ($\alpha = \frac{a}{log(10)}$); circles show the locations of the 21,448 earthquakes (M≥ 3) used; colors corresponding to each earthquake represent the ensemble estimate of $\mu, K$ and $\alpha$.

**Figure 3:** Estimates of the spatially invariant parameters **(a)** $c \: [days]$ **(b)** $\omega$ **(c)** $d \: [km^2]$ **(d)** $\rho$ and **(e)** $\gamma$, corresponding to all 96,000 solutions, as a function of the number of Voronoi cells used, are shown using empty circles; the median estimate corresponding to each Voronoi complexity level is shown by the continuous red line; the continuous grey lines corresponds to the 95% confidence interval of the estimates as a function of complexity level; the horizontal continuous and dashed magenta lines show the value of the ensemble estimate and corresponding 95% confidence interval respectively.

**Figure 4: (a-c)** Spatial variation of the **(a)** magnitude of completeness ($M_c^1$) estimated using the method proposed by Clauset et al. [2009] **(b)** magnitude of completeness ($M_c^2$) estimated using the method proposed by Wyss and Wiemer [2000] **(c)** b-value ($b_{val}$); All earthquakes with M≥0 are used for the estimation of these maps; the $b_{val}$ map assumes that $M_c^1$ is the true completeness magnitude; the maps are only shown at the location of 21,448 (M≥3) earthquakes in the studied region; colors corresponding to each earthquake represent the ensemble estimate of $M_c^1$, $b_{val}$ and $M_c^2$; **(d)** correlation between $M_c^2$ and $M_c^1$; the filled black circles indicate the median $M_c^2$ in a given bin of $M_c^1$; grey bars show the 95% confidence interval for each value; the continuous blue line corresponds to the equation $M_c^2 = M_c^1$.

**Figure 5:** Correlation between the median values (black circles) and 95% confidence interval (grey bars) of **(panels a and c)** $\alpha = \frac{a}{\log 10}$ with the median value of $K$ and $\mu$ respectively, **(panel b)** K with median value of $\mu$ and **(panels d-f)** $b_{val}$ with of the median value of $\mu$, $K$ and $\alpha$ respectively.

**Figure 6: (Panel a)** Spatial variation of the branching ratio ($n$); **(Panel b)** difference $\alpha - b_{val}$ between the exponent $\alpha = \frac{a}{log(10)}$ of the aftershock productivity and the b-value; circles show the locations of the 21,448 earthquakes (M≥ 3) used in this study; the color of each earthquake represents respectively the ensemble estimate of $n$ and $\alpha - b_{val}$ at the location of the earthquake.

**Figure 7:** Correlation between $n$ **(panel a)**, $\alpha$ **(panel b)**, $log_{10}(K)$ **(panel c)**, $log_{10}(\mu)$ **(panel d)** and $b_{val}$ **(panel d)** and $log_{10}[Heat\ Flow\ (mW/m^2)]$; Black circles are the median values and the vertical bars delineate the 95% confidence interval of each of these parameters conditioned on the median value of $log_{10}[Heat\ Flow\ (mW/m^2)]$.

**Figure 8:** Correlation of $R_\mu = \mu_{3D} - \mu_{2D}$ **(panel a)**, $R_n = n_{3D} - n_{2D}$ **(panel b)**, $log_{10}(\mu_{3D})$ **(panel c)** and $n_{3D}$ **(panel d)** with hypocentral depths (km); Black circles are the median values and the grey vertical bars delineate the 95% confidence interval

of each of these parameters conditioned on the median value of hypocentral depths (km).

**Figure 9:** Number of Background Earthquakes as a function of the number of Voronoi cells used to partition the study region; Black circles are the median values and the grey bars delineate the 95% confidence of the number of background earthquakes identified for a given number of Voronoi cells.

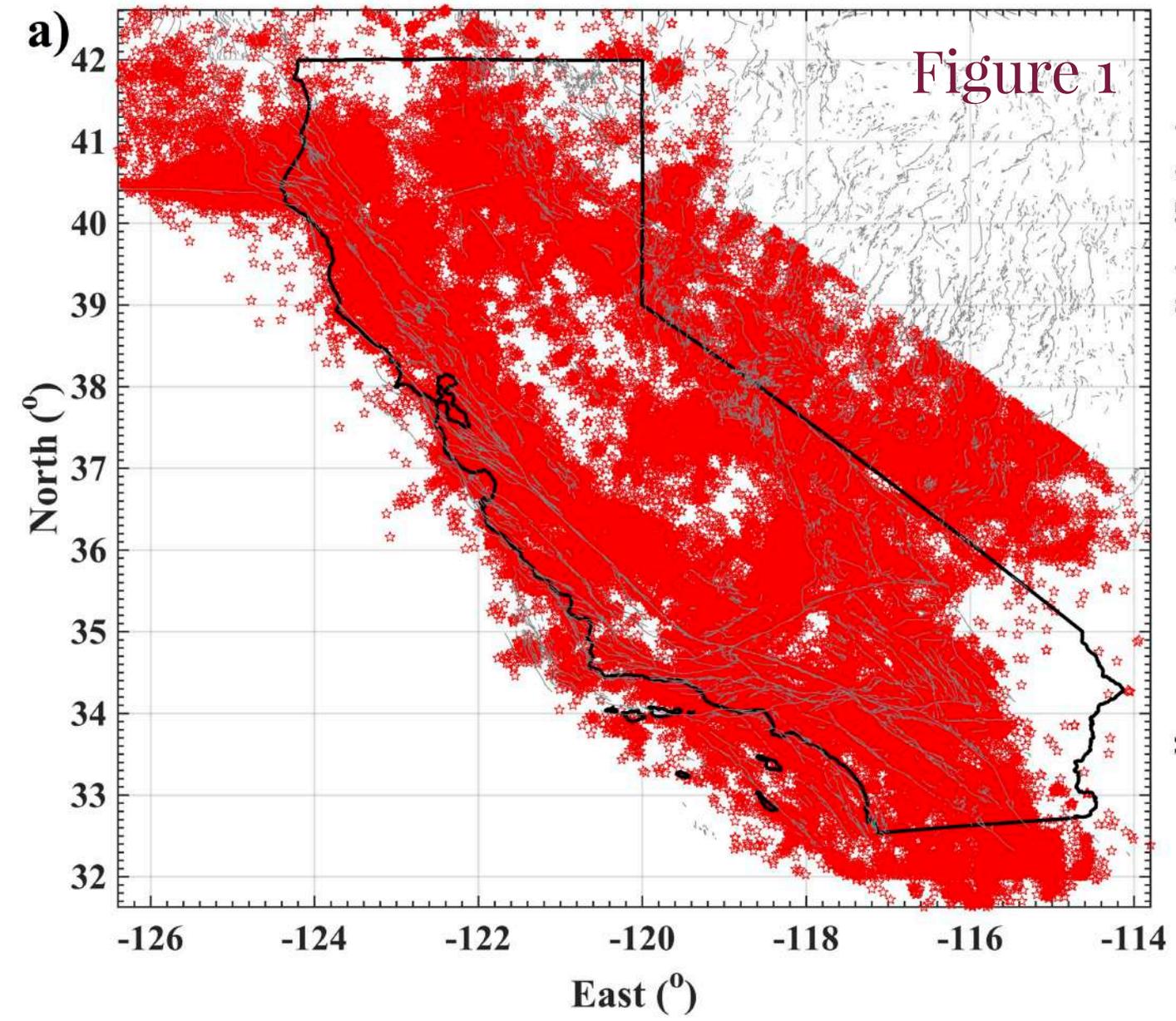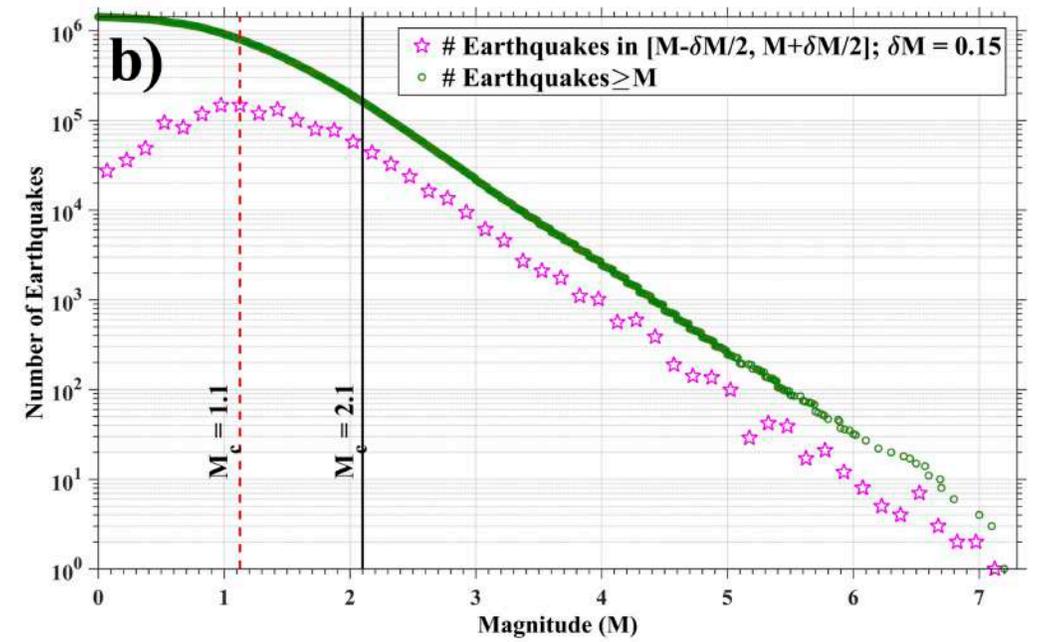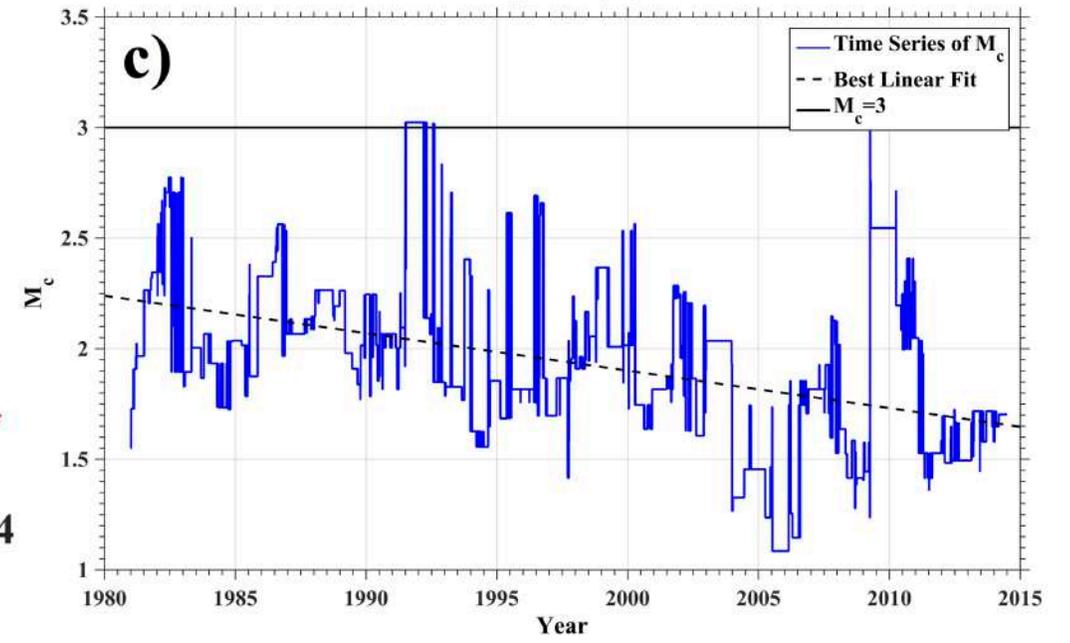

Figure 1

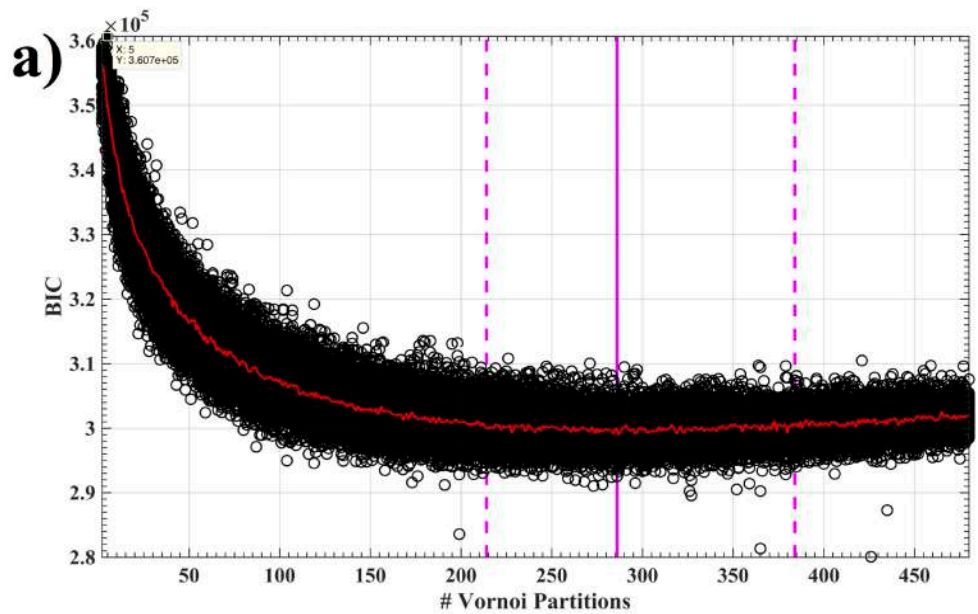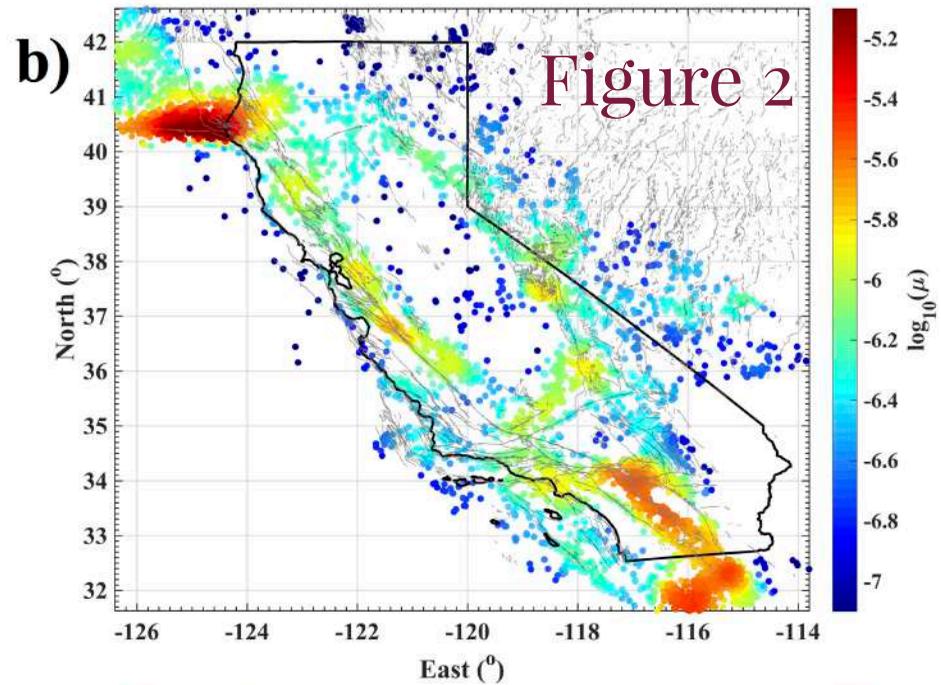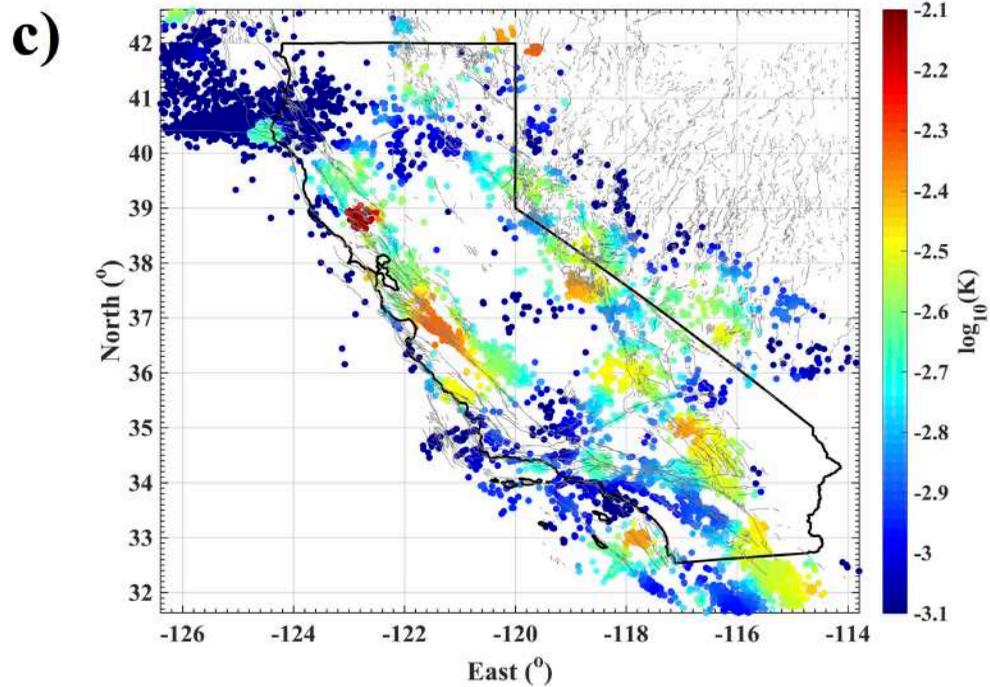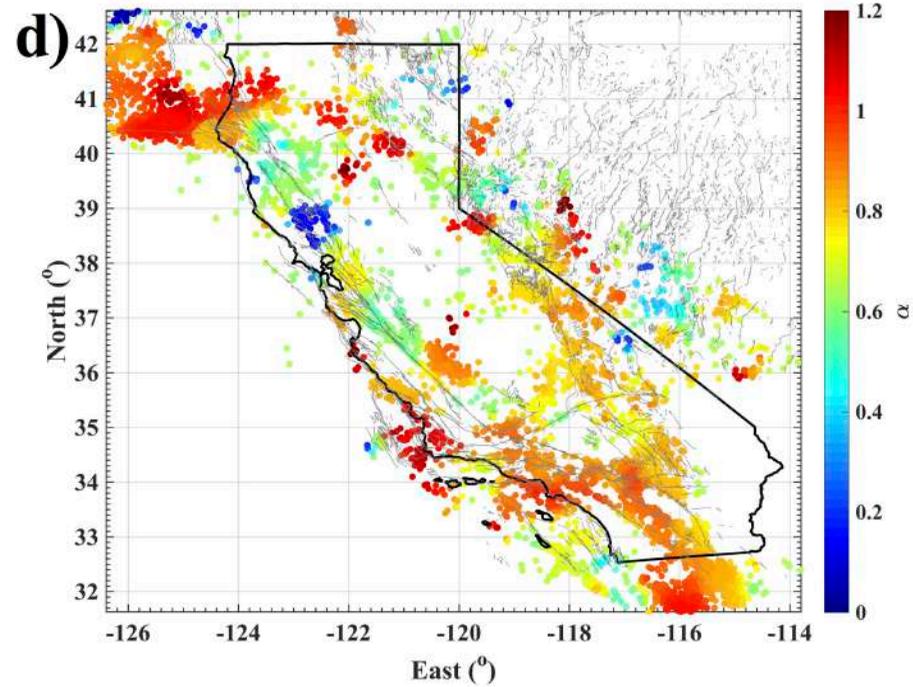

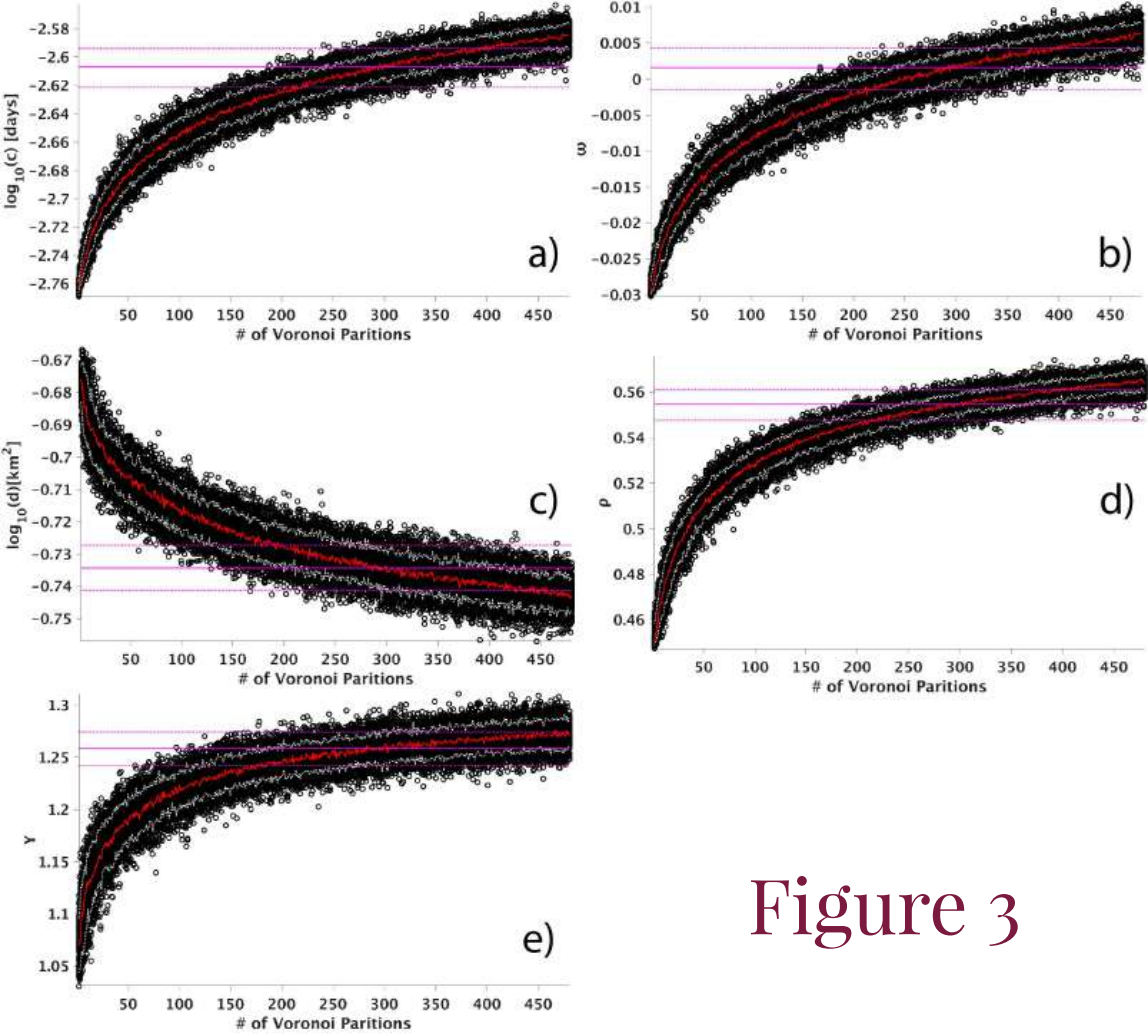

Figure 3

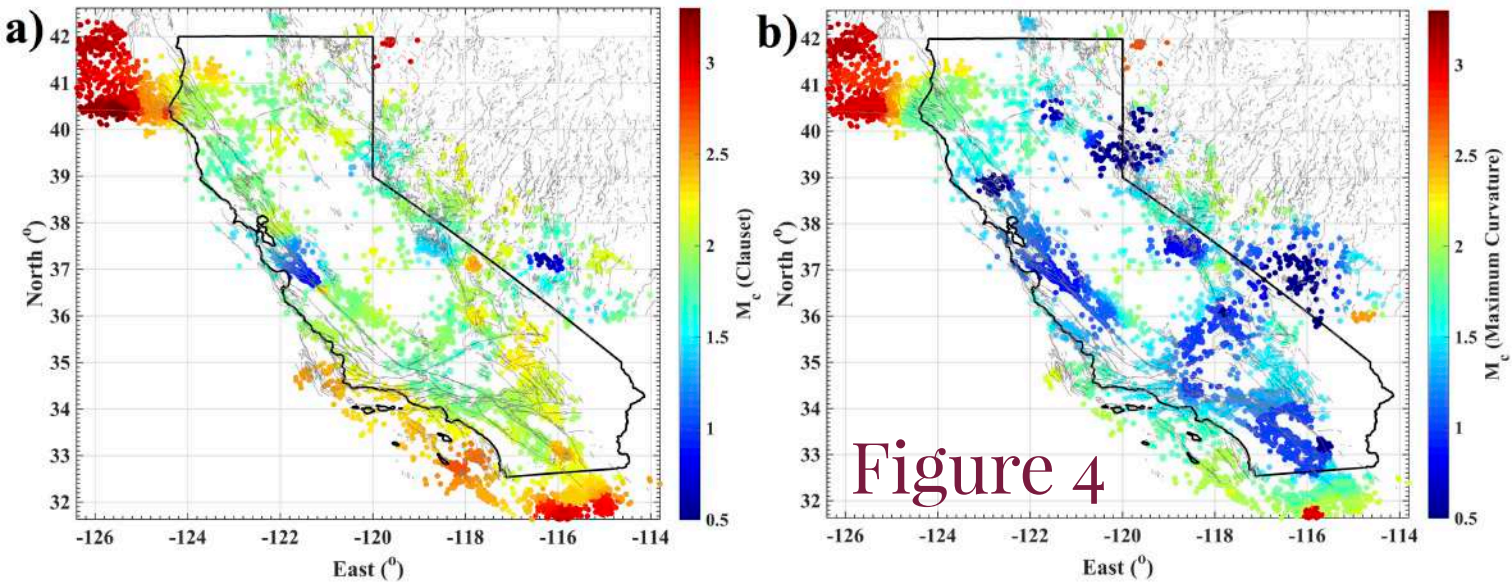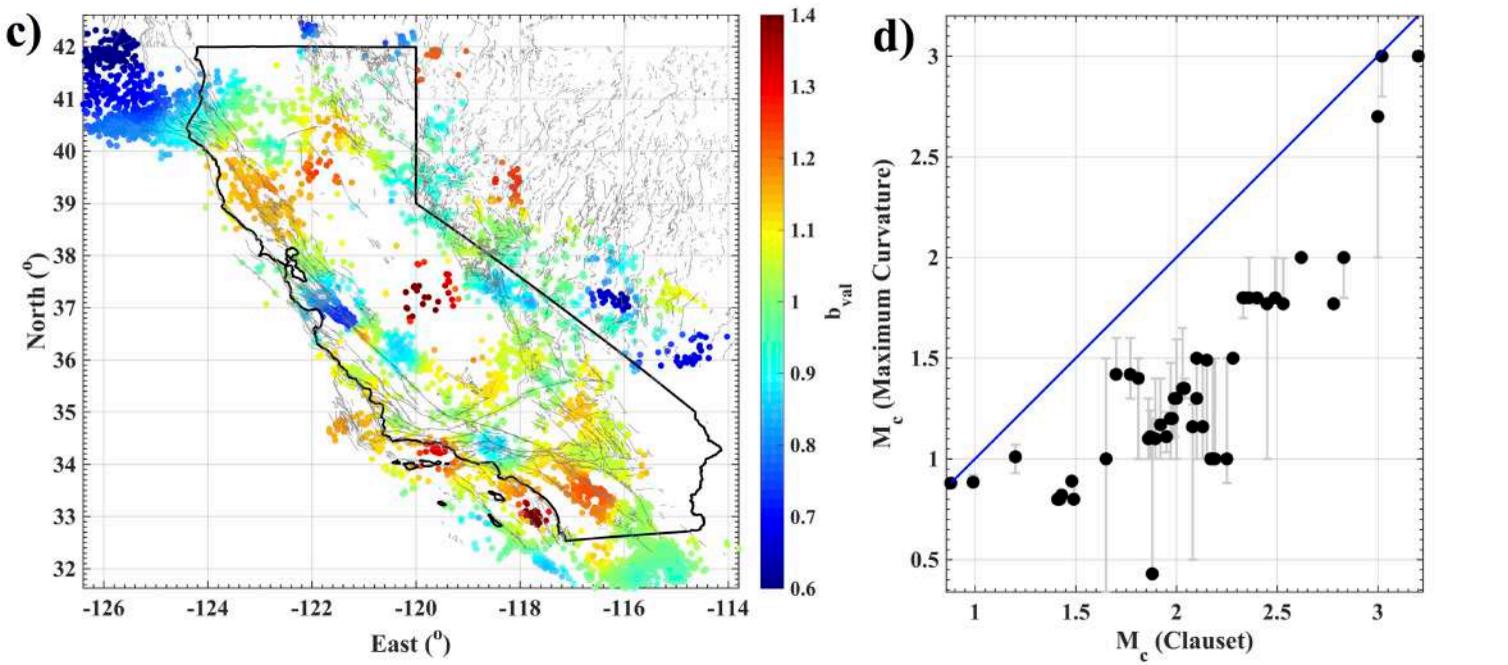

Figure 4

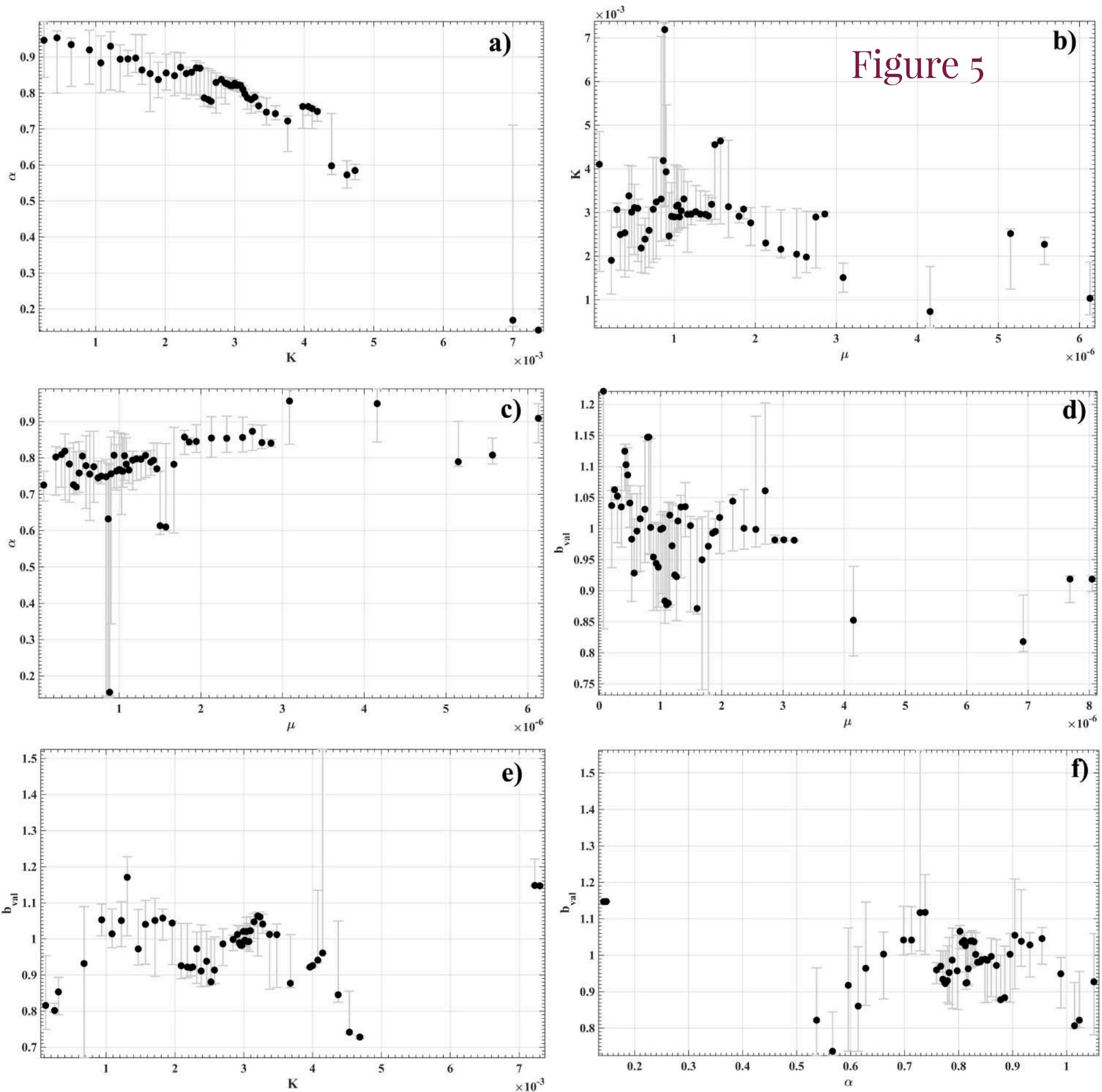

Figure 5

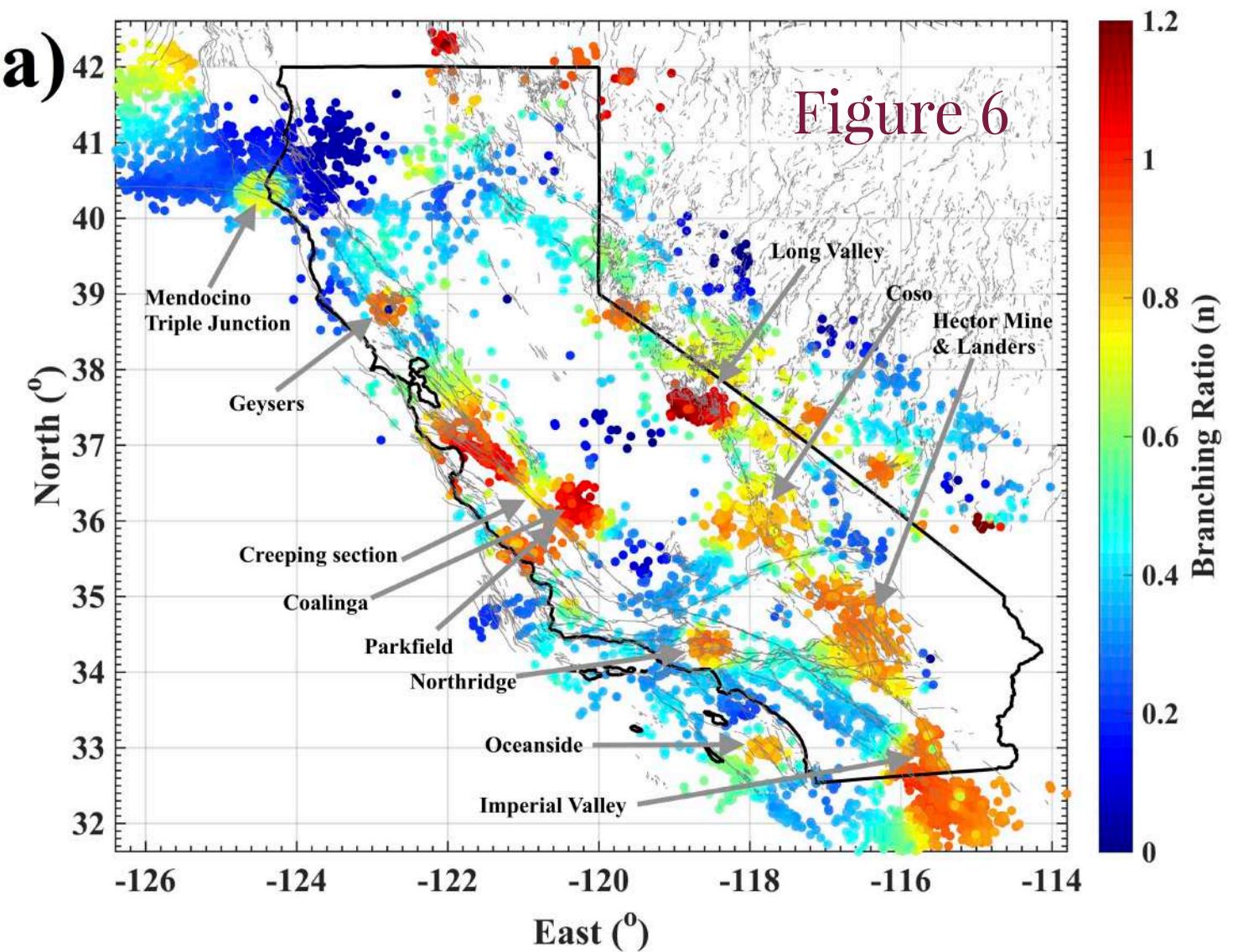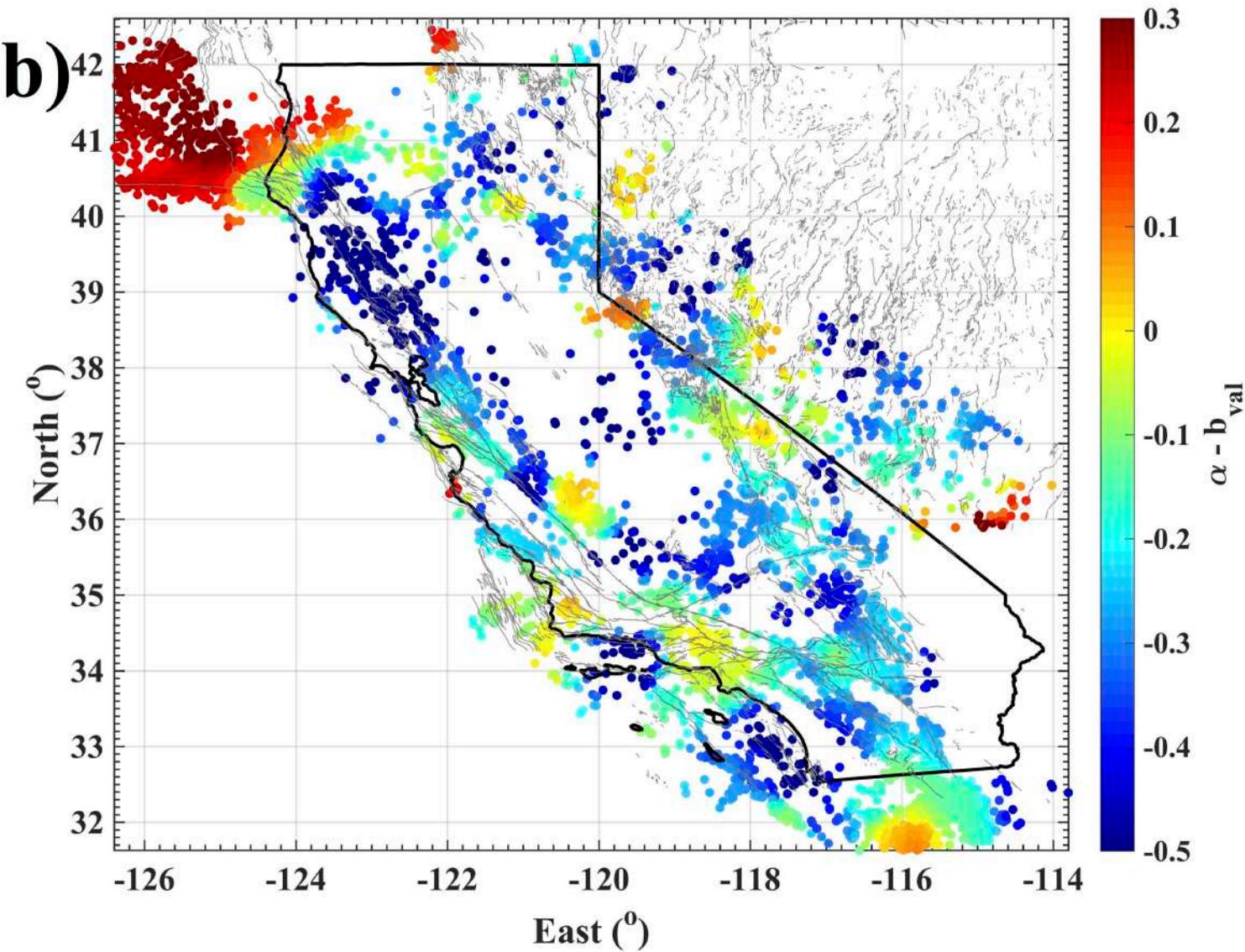

Figure 6

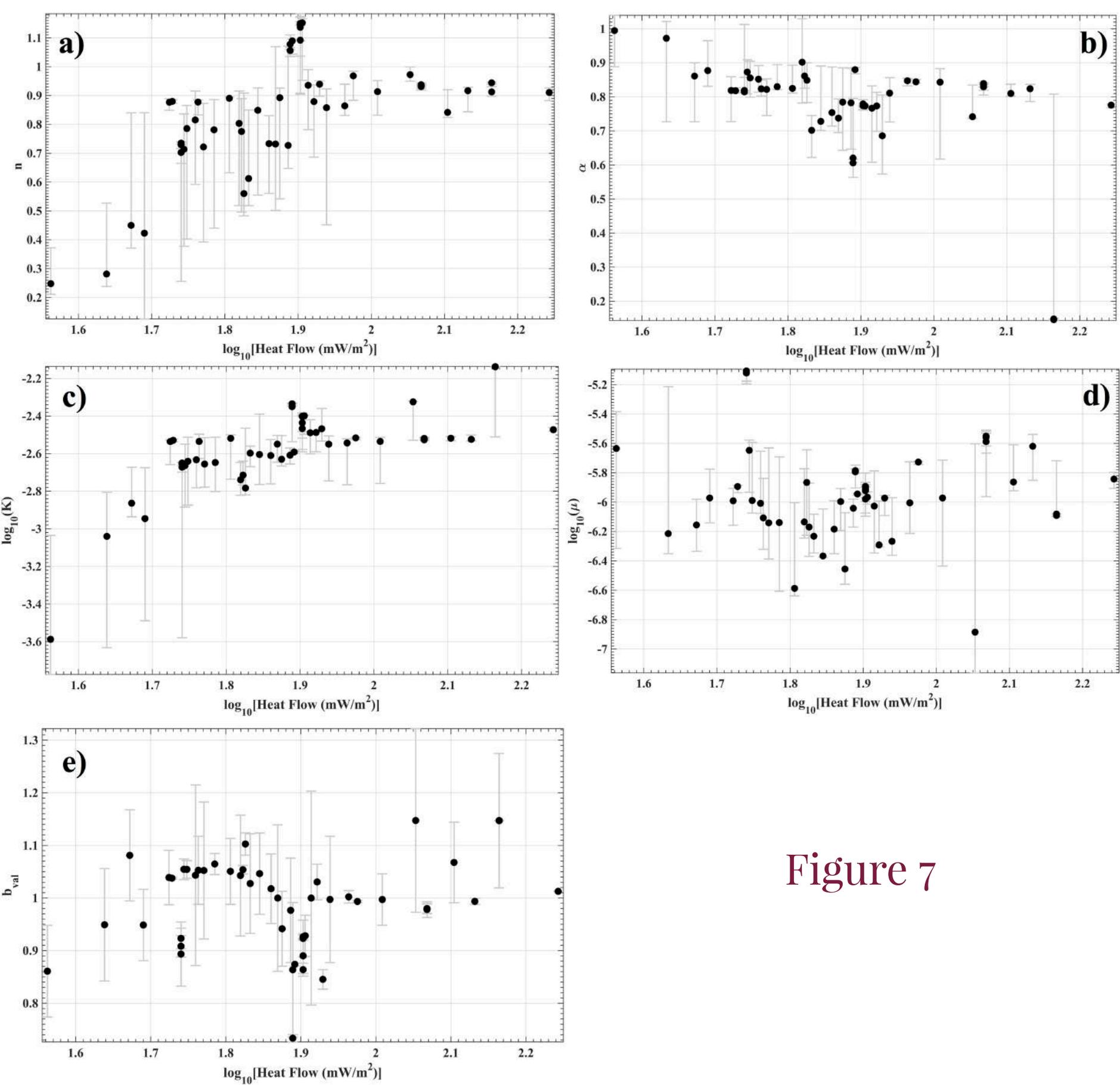

Figure 7

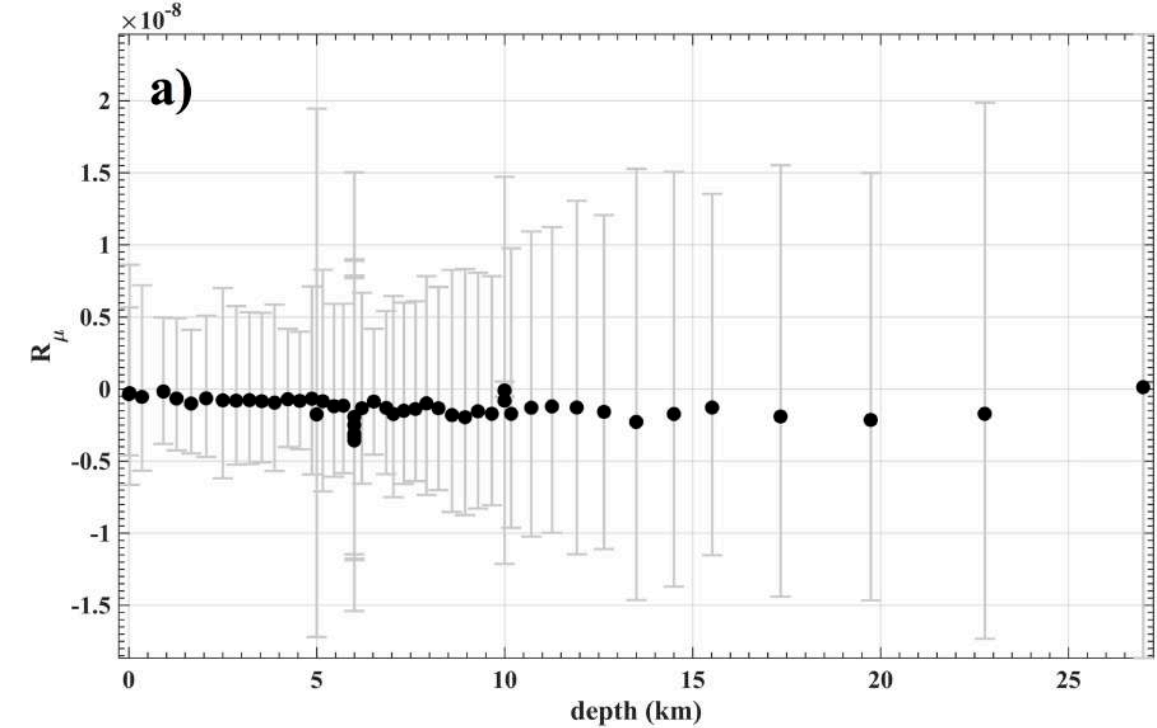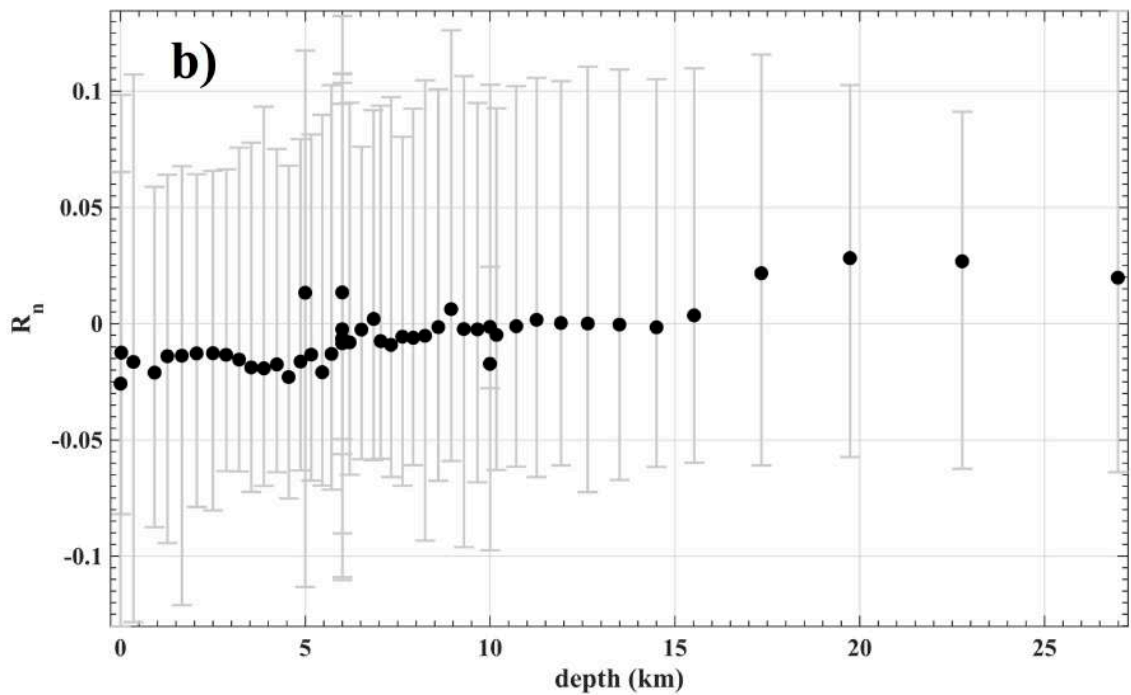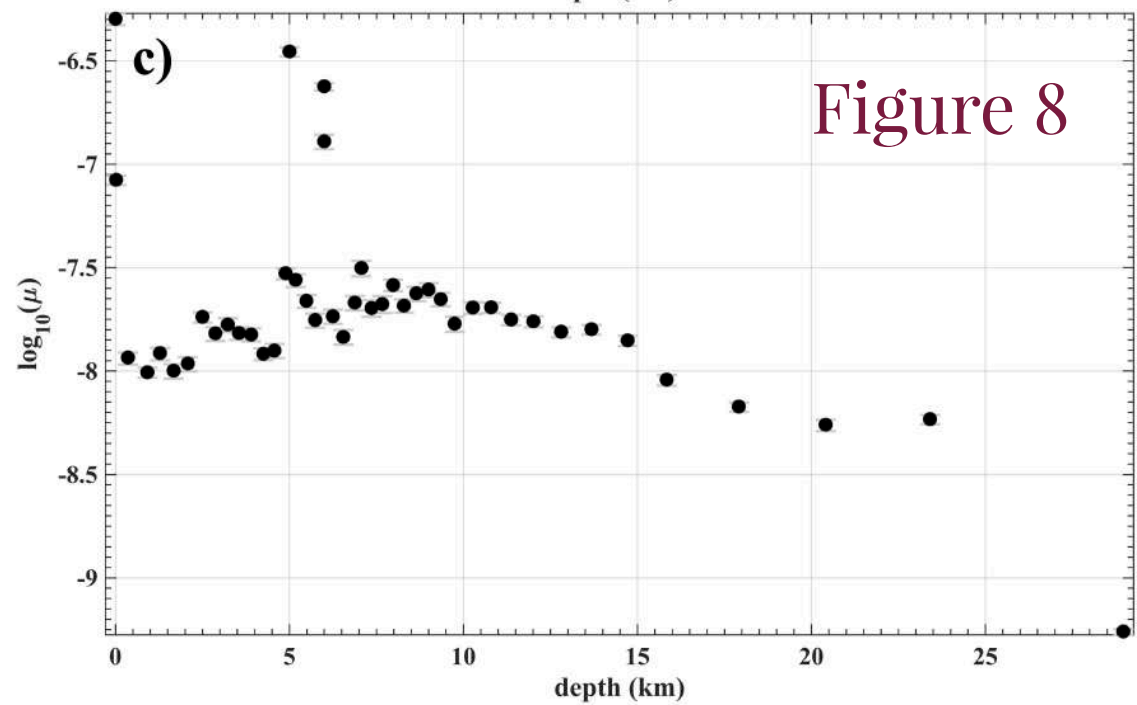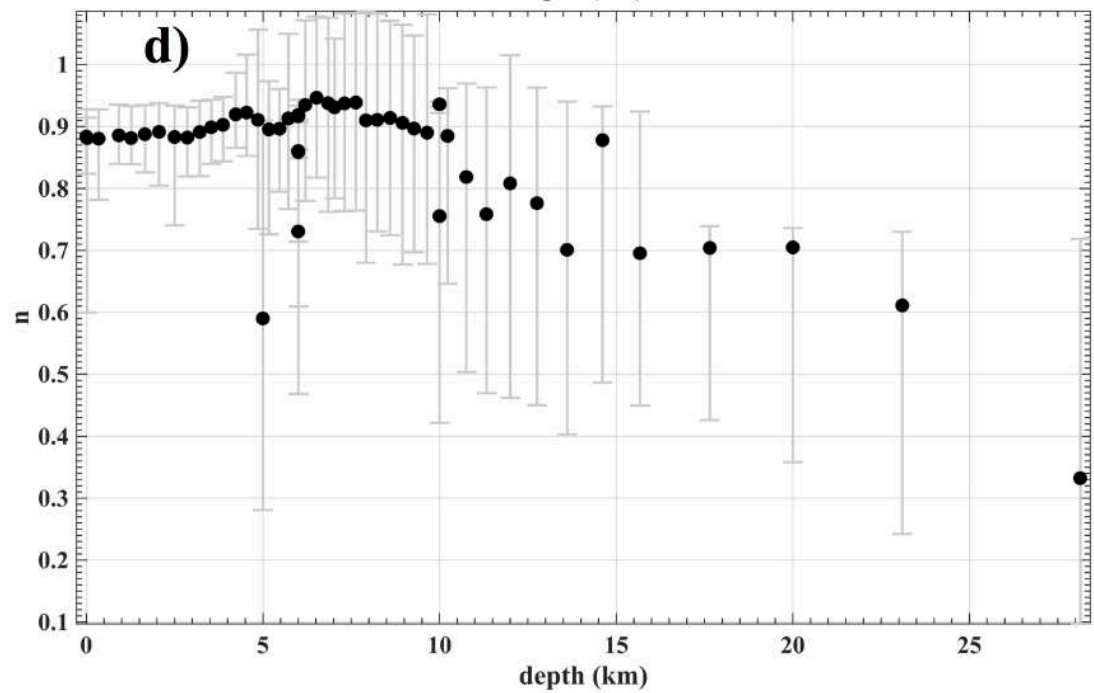

Figure 8

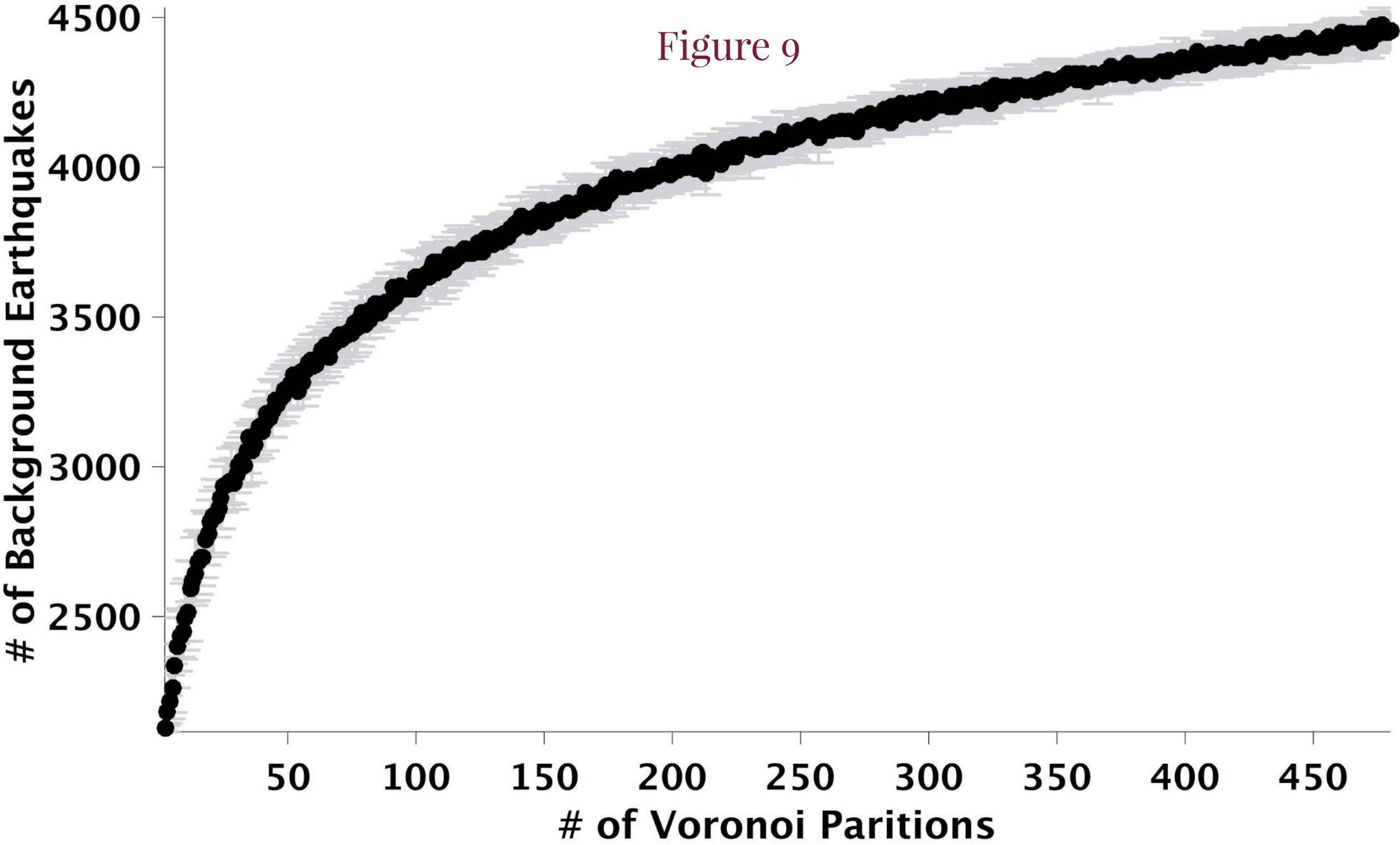

Figure 9

# Estimating Spatially Variable Parameters of Epidemic Type Aftershock Sequence (ETAS) Model with application to California

Shyam Nandan[1], Guy Ouillon[2], Stefan Wiemer[1] and Didier Sornette[3]

**Affiliations:**

[1]ETH Zürich, Swiss Seismological Service, Sonneggstrasse 5, 8092 Zürich, Switzerland

[2]Lithophyse, 4 rue de l'Ancien Sénat, 06300 Nice, France

[3]ETH Zürich, Department of Management, Technology and Economics, Scheuchzerstrasse 7, 8092 Zürich, Switzerland

**Corresponding Author:**

Shyam Nandan, ETH Zürich, Swiss Seismological Service, Sonneggstrasse 5, 8092 Zürich, Switzerland. (shyam4iiser@gmail.com)

**Contents of this file**

> Text S1 to S4
> Figures S1 to S12
> Tables S1 to S3

**Text S1 Synthetic tests**

In this section, we apply the method developed in the section 2.3 to test if 1) the method introduces spurious spatial variations in the parameters even if they are spatially invariant and 2) the method is able to capture the correct patterns of spatial variability in the parameters when they exist.

**Text S1.1 Does the proposed method introduce spurious spatial variations even if the parameters are spatially invariant?**

For testing if the method introduces spurious spatial variations in the parameters even if they are spatially invariant, we first generate a synthetic ETAS catalog with spatially invariant parameters. We first assign a fixed value to each of the 8 parameters ($\boldsymbol{\theta} = \{\mu, K, a, c, \omega, d, \gamma, \rho\}$) (see Table S1). We then simulate the earthquakes using these parameters and the simulation algorithm proposed by Zhuang et al. [2004] over the same spatio-temporal domain as the real catalog. Note that we set the $M_0$ (minimum magnitude below which earthquakes do not trigger aftershocks), $M_{min}$ (magnitude of the smallest possible earthquake), $M_{max}$ (magnitude of the largest possible earthquake) and $b_{val}$ (exponent of the Gutenberg Richter law) values to, respectively, 3, 3, 8.5 and 0.95 in accordance with our assumptions and observations on the real catalog. In Figure S1, we show the spatial distribution of earthquakes generated by the ETAS simulator. We then apply the method proposed in section 2.3 to invert spatially variable $\boldsymbol{\mu}, \boldsymbol{K}$ and $\boldsymbol{a}$ along with spatially invariant $c, \omega, d, \gamma$ and $\rho$ parameters.

In Figure S2, we show the BIC corresponding to all (5000) solutions as a function of the number of Voronoi cells used. We find that the minimum median BIC corresponds to inverted models with only 1 voronoi partition indicating that our method is able to correctly detect the complexity of the input model used to generate the synthetic catalog. Moreover, we also find, using Wilcoxon Ranksum test, that the minimum median BIC (corresponding to only 1 voronoi partition) is significantly smaller than the median BIC corresponding to all other complexity levels. As a result, we obtain spatially invariant estimates of of $\boldsymbol{\mu}$, $\boldsymbol{K}$ and $\boldsymbol{\alpha}$. We report the spatially invariant estimates of all the parameters in Table S1. Comparing the inverted values of the parameters to the input values, we find that that our method correctly estimates the input values of the parameters.

These results demonstrate the ability of our method to not only correctly infer the underlying complexity of the input model used to generate a synthetic catalog, but also the correct values of the input parameters.

**Text S1.2 Is the proposed method able to capture the underlying spatial variation of the parameters?**

For testing if the method is able to capture the correct patterns of spatial variability in the parameters, we first generate a synthetic ETAS catalog using the estimates of parameters ($\boldsymbol{\Theta}, \boldsymbol{\mu}, \boldsymbol{K}$ and $\boldsymbol{a}$) for the real catalog, that covers the same spatio-temporal domain as the real catalog. We make these choices for the generation of the synthetic catalog for it to resemble the real catalog as closely as possible. We then follow the following algorithm:

1. We generate the background earthquakes. To do so, we use the independence probabilities ($IP_j = 1 - \sum_{i=1}^{j-1} \hat{P}_{i,j}$ where $\hat{P}_{i,j}$ is the probability that the $j^{th}$ earthquake has been triggered by the $i^{th}$ earthquake given the estimated parameters $\hat{\Theta}, \hat{\mu}, \hat{K}$ and $\hat{a}$, and is computed using Equation 6), time, location and magnitude of the earthquakes from the real catalog. We compare the independence probability, $IP_j$, of each earthquake, $E_j$, to a random number generated uniformly between 0 and 1. If $IP_j$ is larger than the random number, the earthquake is considered as a background event whose time, location and magnitude is the same as in the real catalog. If not, the event is discarded. This is a semi-stochastic way to simulate the background earthquakes for the synthetic catalog. The advantage of this approach over the conventional strategy to simulate non-homogenous stationary space-time Poisson process [Zhuang et al., 2004; Daley and Vere-Jones, 2002, section 7.4] is that the former allows us to use the location and magnitude of real earthquakes, which possibly capture the geometry of the underlying fault network. We then consider each background earthquake as a parent earthquake.

2. We then assign a set of productivity parameters $K_j$ and $a_j$ to each parent earthquake depending upon the value of the ensemble estimates of the parameters $\hat{K}$ and $\hat{a}$ from the real catalog at their location.

3. For each parent earthquake with magnitude $m_j$, we generate $N_j$ offsprings earthquakes above magnitude $M_0$, where $N_j$ is a discrete Poisson random variable with mean $G_j$ (see Equation 3 for the definition of $G_j$ in Section 2.1). The times and locations of each offsprings earthquakes are simulated stochastically using an Omori kernel in time $\{t - t_j + \hat{c}\}^{-1-\hat{\omega}}$ and a spatial density kernel $\{(x - x_j)^2 + (y - y_j)^2 + \hat{d} * e^{\hat{\gamma}*(m_j-M_0)}\}^{-1-\hat{\rho}}$ respectively, where $\widehat{\Theta} = \{\hat{c}, \hat{\omega}, \hat{d}, \hat{\gamma}, \hat{\rho}\}$ are the previously estimated spatially invariant parameters. We simulate the magnitudes of the offsprings earthquakes using the pdf of a Gutenberg-Richter law with $M_{min} = 3$ as the lower magnitude cutoff, $M_{max} = 8.5$ as the upper magnitude cutoff and a global b-value of 0.95 (estimated for the real catalog, see section 3). We also assume that only the earthquakes above magnitude, $M_0 \geq 3$ are able to trigger aftershocks. We then consider the offsprings earthquakes as the parent earthquakes for the next generation.

4. We repeat steps 2 and 3 until no newer offsprings earthquake is generated.

Note from Figure 6a that the estimated branching ratio for the real catalog sometimes locally exceeds 1, which can lead to an explosive generation of earthquakes. To account for this, we modify the $K_j$ value assigned to the earthquakes (in step 3) for which the local branching ratio exceeds 1 such that the newly assigned branching ratio is equal to 1. This leads to the generation of a non-explosive catalog.

Figure S3 shows the spatial distribution of the earthquakes generated using the proposed algorithm.

We then apply the method proposed in section 2.3 to invert the spatially variable $\boldsymbol{\mu}, \boldsymbol{K}$ and $\boldsymbol{a}$ along with spatially invariant $c, \omega, d, \gamma$ and $\rho$.

In Figure S4, we show the BIC corresponding to all (120,000; 200 for each given number partition) solutions as a function of the number of Voronoi cells used. We find that the minimum in the median BIC corresponds to 284 voronoi partitions (indicated using solid magenta line in Figure S3). As proposed in the section 2.3, we find the complexity range, [178- 428], in which the median BIC corresponding to each complexity level is not significantly different from the minimum median BIC. Note that both the optimal complexity level and complexity range, identified by our method for the synthetic catalog, nearly coincide with the optimal complexity level (286) and the complexity range, [214-384], observed in the case of the real catalog (see Figure 2a). Since the parameters inverted from the real catalog are used as the input parameters for the generation of the synthetic catalog, the near coincidence of the optimal complexity level and the complexity range for the synthetic catalog and the real catalog indicate that our method correctly detects the complexity of the underlying model.

In Figure S5, we show the spatially varying estimates of $\boldsymbol{\mu}$, $\boldsymbol{K}$ and $\boldsymbol{\alpha} = \frac{a}{log10}$ at the location of the earthquakes present in the synthetic catalog. The left panels in the figure show the spatial variation of the input parameters (used to generate the synthetic catalog) while the right panels show the spatial variation of the same parameters inverted from the synthetic catalog. Visually comparing the input and inverted parameters, we find that our method is quite successful in inverting the underlying spatial patterns of the three spatially varying input parameters. In addition to that, the inverted values of the spatially invariant parameters ($c, \omega, d, \gamma$ and $\rho$), shown in Table S2, are very close to the input values (see the solid magenta lines in Figure 3a-e).

It is also important to note that, even though the parameters $\boldsymbol{K}$ and $\boldsymbol{\alpha}$ are correlated with each other (see Figure 5a) and can compensate for each other during the inversion process, our method correctly detects the input patterns of both these parameters (compare Figure S5c and S5e to S5d and S5f respectively).

In Figure S6a-c, we compare the input and inverted values of $\boldsymbol{\mu}, \boldsymbol{K}$ and $\boldsymbol{\alpha}$ more quantitatively. For clarity, we divide the inverted values of each of the three parameters, which are estimated at the location of earthquakes in the synthetic catalog (see right panels in Figure S5), separately in 50 bins conditioned on the values of the corresponding input parameters. We then plot the median value and 95% CI of each parameter versus the corresponding median value of the input parameter, obtained within each of the 50 bins. We find that the inverted values of $\boldsymbol{\mu}, \boldsymbol{K}$ and $\boldsymbol{\alpha}$ are highly correlated with the corresponding input values, with correlation coefficient of 0.98, 0.92 and 0.95 respectively. However, we do find that the inverted versus the input values of these parameters systematically deviate from the x=y line shown in each of the figures, especially in the case of $\boldsymbol{\mu}$. The parameter $\boldsymbol{\mu}$ seems to be underestimated in the regions of high background seismicity rate. The underestimation of high values of $\boldsymbol{\mu}$ can be rationalized if we consider that a high background seismicity rate leads to a high density of earthquakes in a given region, which can appear as clustered. This can further lead to the misclassification of some background earthquakes in regions with very high background seismicity rate as aftershocks. As a consequence, the background rates in the regions of high background seismicity rate will be underestimated.

Nevertheless, the high degree of correlation between the inverted and input parameters as well as the similarity of spatial patterns in the map of inverted and input $\mu$, $K$ and $a$ (see Figure S5a-c) indicate that our method is capable to capture the correct patterns of spatial variability in these parameters.

**Text S2 Goodness of fit of Omori kernel with fixed c-value to observed aftershock decay rate**

In our formulation of the ETAS model, we have assumed the c-value of the Omori kernel to be independent of the magnitude of the mainshock while it might actually depend on it, either due to physical reasons [Dieterich, 1994; Narteau et al., 2002] or due to short term aftershock incompleteness [Hainzl, 2016; Helmstetter et al., 2006]. As a result, the Omori kernel with fixed c-value might not appropriately describe the decay rate of aftershocks in the real catalog. In the following, however, we demonstrate with the goodness of fit test proposed by Clauset et al. [2009] that an Omori kernel with fixed c-value fits the observed decay rate of aftershocks very well.

To test the goodness of fit of the Omori kernel with fixed c-value, we first extract the empirical decay rate of aftershocks from the branching structure of the catalog, obtained after the calibration of the ETAS model. Note that the branching structure quantifies the probability ($P_{ij}$) that the j[th] earthquake (occurring at time $t_j$) has been triggered by the i[th] earthquake occurring at time $t_i$, where $t_j - t_i > 0$. Given this probability matrix, we can extract from the branching structure direct mainshock-aftershock pairs and their corresponding time differences ($\Delta t_{ij}$). This is simply done by generating a uniform random number between 0 and 1 and comparing it to $P_{ij}$. If $P_{ij}$ is greater than the random number, then the corresponding $\Delta t_{ij}$ is chosen. In this manner, we were able to extract ~17,000 values of $\Delta t_{ij}$. Indeed, this extraction procedure yields only those mainshock-aftershock pairs that are directly related to each other. Using these $\Delta t_{ij}$'s, we then estimate their empirical probability density function, which integrates to 1 within the minimum and maximum time interval in the real catalog. This empirical PDF is shown using blue crosses in Figure S7.

Next, for each of the selected mainshocks ($t_i$), we simulate as many aftershocks as there are in the real catalog. The times of these aftershocks are simulated using the Omori kernel with parameters p=1.0051 and $c = 10^{-2.59}$ days. Note that these parameters were obtained from the calibration of the spatially variable ETAS model on the real catalog. We make sure that the Omori kernel integrates to 1 in the time period between the occurrence of the mainshock and the end time of the catalog. By doing so, we ensure that the mainshock would have as many simulated aftershocks as there are in the real catalog. Finally, we compute the PDF of the $\Delta t_{ij} = t_j - t_i$ for all the mainshock-aftershock pair in the simulated catalog within the same time limits of $10^{-4}$ and $10^{4.2}$ days. This empirical PDF is shown using solid red line in Figure R1.

We can clearly see that the empirical PDF simulated using the Omori kernel with fixed c value fits well the empirical PDF obtained from the real catalog.

To further quantify the goodness of fit, we use the standard goodness of fit test proposed by Clauset et al. [2009] (section 4.1) which is composed of the following steps:

1. We compute the Kolmogorov Smirnov (KS) distance between the best fit Omori kernel (with parameters p=1.0051 and $c = 10^{-2.59}$ days) and the empirical aftershock decay rate obtained from the real catalog ($KS_{real}$). Note that these parameters are obtained by calibration of the ETAS model.
2. Using the best Omori kernel fit, we simulate aftershock sequences for each mainshock present in the real catalog. We keep the number of aftershock for each mainshock to be the same as in the real catalog. As before, we ensure that the Omori kernel, for each mainshock, integrates to 1 in the time period between the occurrence of the mainshock and the end time of the catalog. By doing so, we ensure not only that the mainshock would have as many simulated aftershocks as there are in the real catalog, but also that the simulated catalog would exhibit the same finite size effects as does the real catalog.
3. For each of the simulated catalogs, we re-estimate the parameters of the Omori kernel.
4. We compute the KS distance between the new Omori kernel and the empirical aftershock decay rate obtained from the simulated catalog ($KS_{synthetic}$).
5. We repeat steps 2-4 10,000 times.

The histogram of $KS_{synthetic}$ is shown in Figure S8. For comparison, $KS_{real}$ is shown using a solid grey vertical line. We find that, in nearly 21% of the cases, the $KS_{synthetic}$ is larger than $KS_{real}$. As a result, in accordance with the relatively stringent criteria proposed by Clauset et al. [2009], we can safely conclude that the Omori kernel with a fixed c-value is a reasonable hypothesis for the aftershock decay rate observed in real catalog, which cannot be rejected at all standard statistical significance levels with a p-value of 0.21.

**Text S3 Influence of short term aftershock incompleteness on our results**

In order to assess the influence of short term aftershock incompleteness on our results, we first modify the Omori kernel of the ETAS model to $\frac{1}{\left\{t-t_i+c_0 e^{\eta(m_i-M_0)}\right\}^{1+\omega}}$. Note that the c-value in this modified Omori kernel depends on the magnitude $m_i$ of the mainshock. As long as the parameter $\eta$ of the modified Omori kernel is positive, the c-value would increase with the magnitude of the mainshock, which is consistent with idea of short term aftershock incompleteness. We then calibrate the modified ETAS model (ETAS-mod) on the earthquake catalog ($M \geq 3$) used in this study using the method proposed in section 2.3 in the main text. In Figure S9, we show the penalized log likelihood ($BIC_{mod}$) of the ETAS-mod as a function of the number of voronoi partitions. The red circles show the median value of $BIC_{mod}$ and the error bars show the 95% confidence interval. In the same figure, we also show the penalized log likelihood ($BIC_{unmod}$) corresponding to the unmodified ETAS model (ETAS-unmod) using a solid blue line. It is evident from the figure that $BIC_{mod}$ is not significantly larger than $BIC_{unmod}$. We also verify it using the Wilkoxon Ranksum test.

The similar performances of ETAS-mod and ETAS-unmod in describing the spatio-temporal distribution of earthquakes ($M \geq 3$) in the catalog, in terms of penalized log-likelihood, indicate that both models are equally likely.

We further compare the spatially variable and spatially invariant parameters obtained from ETAS-mod and ETAS-unmod models. In figure S10, we plot the four spatially variable parameters: background seismicity rate ($\mu$); branching ratio (n); pre-factor of the productivity law (K) and exponent of the productivity law ($\alpha$), obtained from the calibration of the ETAS-mod model versus the ones obtained from calibration of the ETAS-unmod model on the catalog. In the figure, we also plot the x=y line (in red) for comparison. We can clearly see that the spatially variable parameters obtained from calibration of both versions of the ETAS model is nearly identical. It automatically implies that the estimates of the parameters $K$ and $\alpha$ obtained from the modified ETAS model are also negatively correlated.

In Table S3, we show the ensemble estimates of the spatially homogenous parameters obtained from both models. Again, we find that the calibration of both models on the catalog yields nearly equivalent spatially invariant parameters.

Finally, we also find that the ensemble estimate of $\eta$ is -0.19. The negative value of $\eta$ indicates that the c value, which is thought to indicate the short term aftershock incompleteness duration, decreases with the magnitude of the mainshock. In fact, this observation is inconsistent with the hypothesis that short term incompleteness increases with the magnitude of mainshock. While this hypothesis might be true, it is not supported by the data when we consider only earthquakes with $M \geq 3$.

Finally, it is interesting to note that several physics-based models such as the stress corrosion model [Scholz, 1968; Narteau et al., 2002] and rate and state model [Dieterich, 1994; Dieterich et al., 2000] postulate that larger amplitudes of stress perturbations can lead to a decrease in duration of the non-power-law regime in the rate of aftershock decay, which would imply that the c value of the Omori law would decrease with the magnitude of the mainshocks (it is assumed that larger earthquakes would cause larger stress perturbations). This hypothesis seems to be in agreement with the negative value of $\eta$ observed in the case of ETAS-mod.

**Text S4 Is the estimated branching ratio correlated with seismicity rate?**

In the following, we show first by comparison of the maps of seismicity rate and branching ratio obtained from the real catalog, and then by controlled synthetic experiments, that the two quantities are not correlated.

**Text S4.1 Comparison of maps of branching ratio and seismicity rate in the real catalog**

In Figure S11, we show the map of the total seismicity rate. In order to compute this map at the location of the 21,448 $M \geq 3$ earthquakes, we adopt the following strategy. For a given spatial Voronoi partitioning scheme used during the calibration procedure, we first obtain the estimate of the average seismicity rate for each of the spatial cells by counting the number of earthquakes enclosed within each of the cells, and by dividing this number by the area of the cells (in $km^2$) and total time period of the catalog (in years). All earthquakes enclosed in each of the spatial cells are then assigned the corresponding estimate of the average seismicity rate. We repeat this two steps procedure for all the 34,200 Voronoi partition schemes corresponding to the selected solutions within the optimal complexity range (shown Figure 2a) to obtain 34,200 individual seismicity rate maps. Finally, we obtain the ensemble seismicity rate map (shown in Figure S11) by weighting all the individual rate maps with weights that are computed according to equation (13).

On comparing this seismicity rate map to the map of the parameters $n$ (shown in Figure 5a), we find that, while there exist regions in the maps that indeed display both a high seismicity rate and a high value of $n$, there exists prominent counter examples to this observation. In Figure S11, we have marked some of the counter examples for easier visualization using a dashed arrow. We find that, indeed, there exist prominent regions, such as offshore Mendocino, where we observe a very high seismicity rate and yet our method inverts a very small value of $n$. Furthermore, we have also indicated some regions in Figure S11 that have overall a low seismicity rate but a very high branching ratio. The existence of regions of both types (1. high seismicity rate and low $n$; 2. low seismicity rate and large $n$) clearly demonstrates that (1) our method does not exclusively identify high $n$ in regions of high seismicity rate and (2) our method also associates regions of high seismicity rate with low $n$.

**Text S4.1 Controlled synthetic experiment**

We design the following experiment. We simulate synthetic earthquake catalogs with varying total seismicity rates but a fixed branching ratio. The total seismicity rate is varied by means of changing the background seismicity rate. For each of the synthetic catalogs, we then apply our estimation procedure and estimate the parameters that were used for simulation. We then investigate the correlation between the estimated branching ratio and total seismicity rate.

In our experiment, we vary the total seismicity rate by scanning the background seismicity rate from $10^{-7}$ to $10^{-6.1}$ earthquakes per day per $km^2$. The two aftershock productivity parameters, the branching ratio (n) and the exponent of the productivity law ($\alpha$), are fixed respectively to values 0.7 and 0.8. The times and the location of the aftershocks are simulated using the Omori kernel ($\frac{1}{(t+c)^p}$) and spatial kernel ($\frac{1}{(x^2+y^2+de^{\gamma m})^q}$). The parameters $\{c, p, d, \gamma, q\}$ of these two kernels are set to values $\{10^{-2.6}, 1.1, 10^{-0.75}, 1.24, 1.6\}$. The magnitudes of the earthquakes are simulated using a Gutenberg-Richter law with exponent $b = 1$ in all the simulations. In our simulations, we also assume that the magnitude threshold above which earthquakes start triggering other earthquakes is $M_0 = 3$ and the magnitude of the largest earthquake than can occur is 8.5. As the ETAS model is highly stochastic, for a given set of parameters $\{\mu, n, \alpha, c, p, d, \gamma, q\}$, we perform numerous (200) simulations. For each of the simulated catalogs, we use our method to estimate the underlying parameters. We then compute the expected total seismicity rate from all the 200 simulated catalogs for a given set of parameters. From the estimated parameters of all the 200 simulated catalogs, we then estimate the median branching ratio and corresponding 95% confidence interval. We repeat this process of simulation and estimation by varying the background seismicity rate from $10^{-7}$ to $10^{-6.1}$ earthquakes per day per $km^2$. In Figure S12, we show the estimated branching ratio and its 95% confidence interval as a function of the total seismicity rate. It is clear from the figure that there is no correlation between the estimated branching ratio and the total seismicity rate. In fact, regardless of the total seismicity rate, the estimated branching ratio is always close to the true branching ratio.

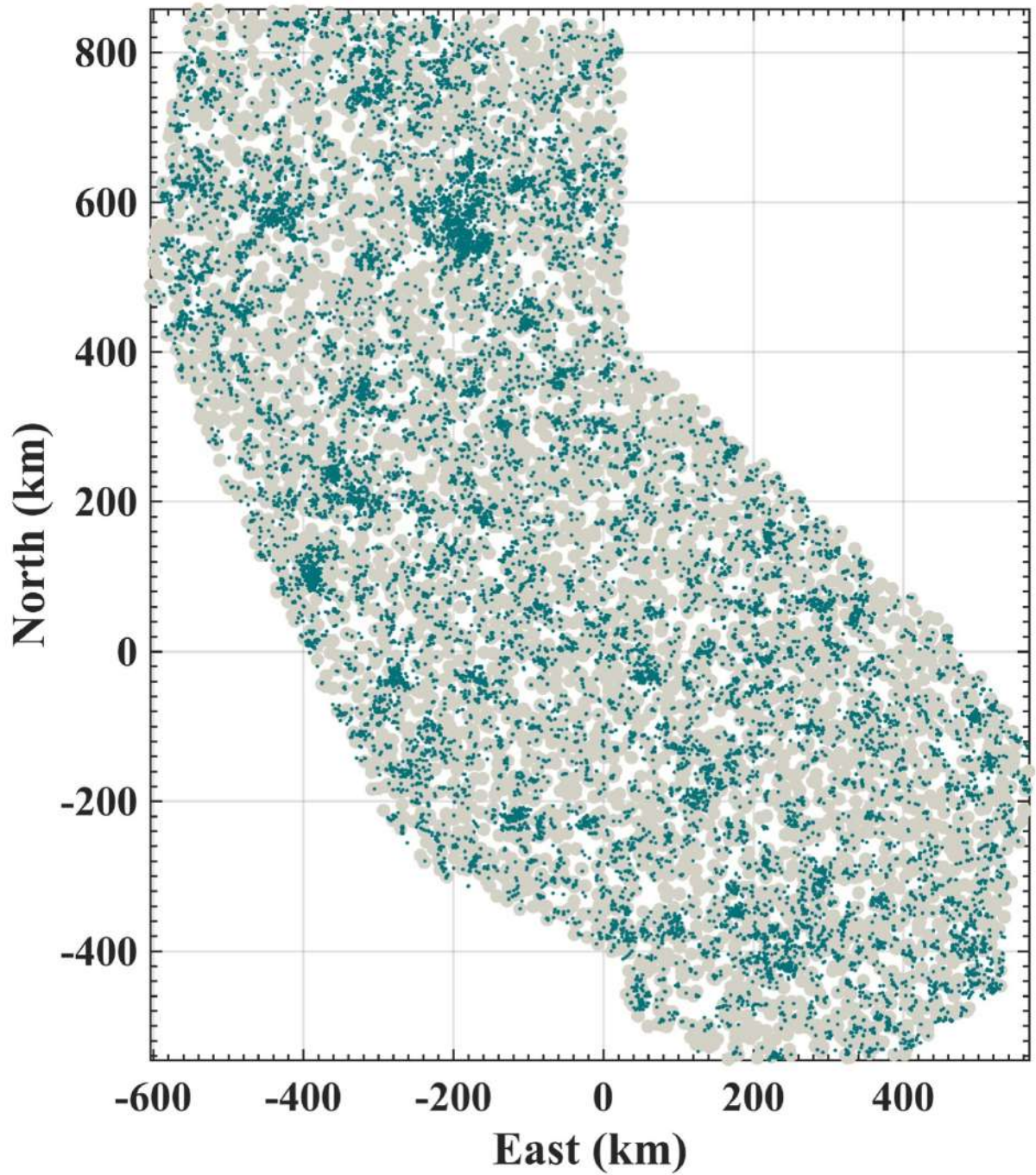

**Figure S1**: Spatial distribution of the earthquakes generated in the synthetic catalog over the same spatio-temporal domain as the natural catalog; grey circles and green dots show the spatial distribution of the background earthquakes and aftershocks respectively.

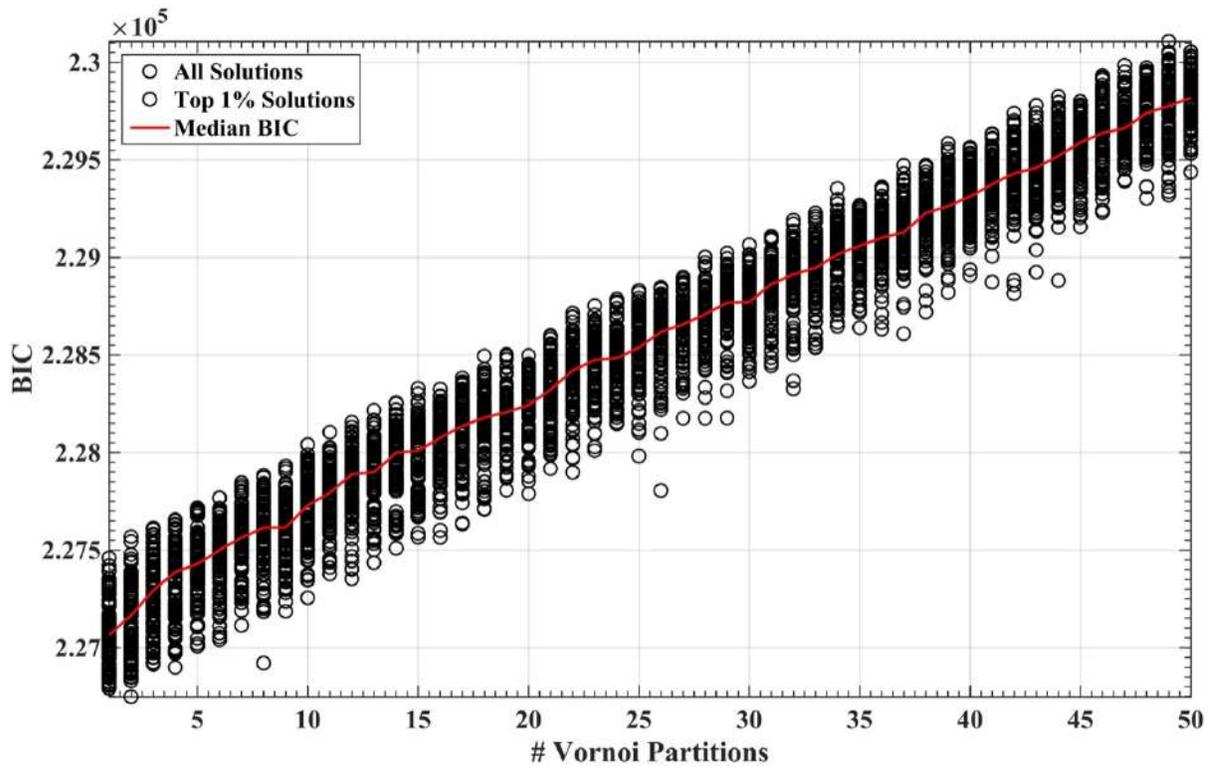

**Figure S2**: The BIC corresponding to 5000 solutions as a function of the number of Voronoi cells used is shown using black circles; the median BIC corresponding to each Voronoi complexity level is shown using a solid red line.

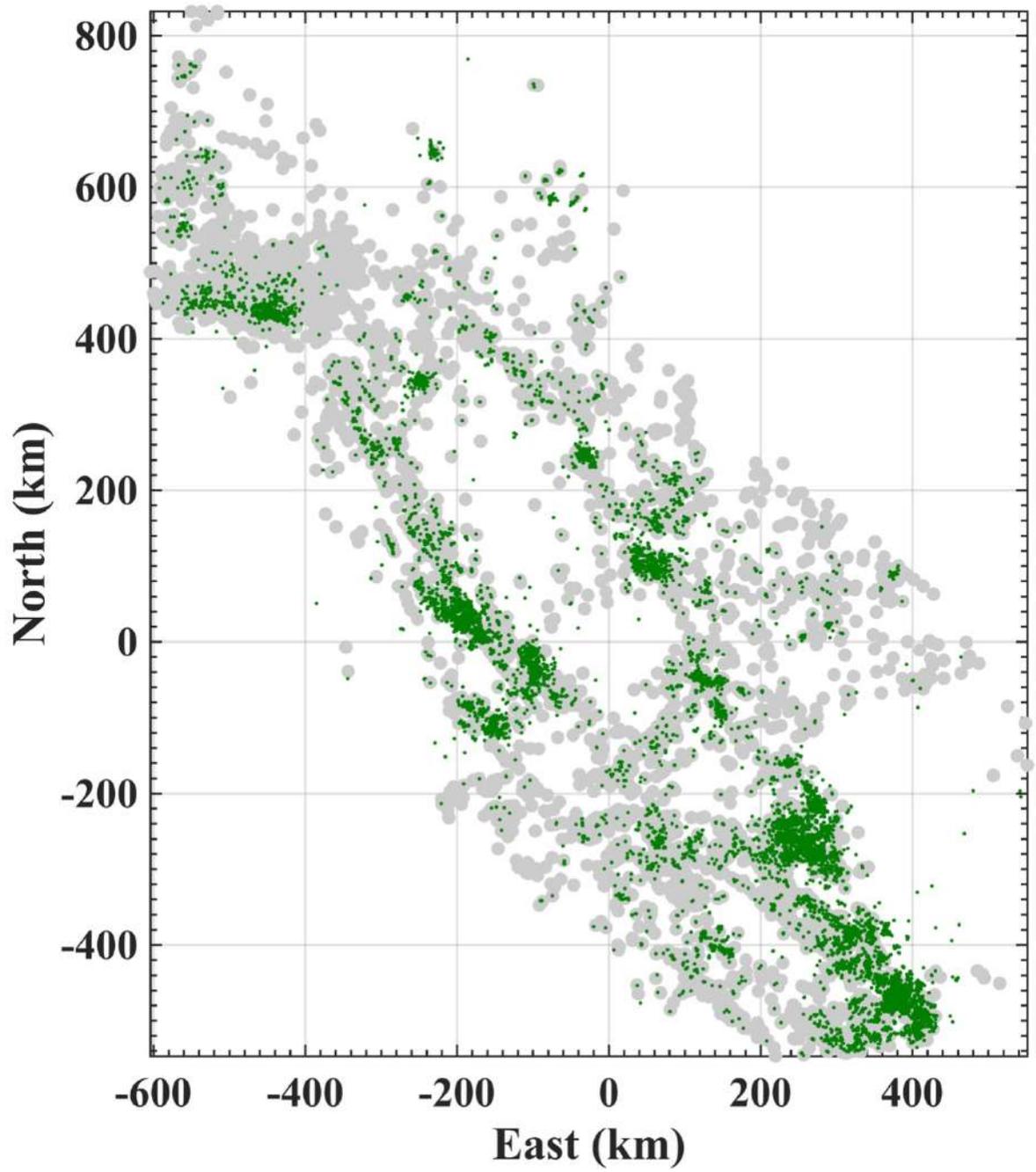

**Figure S3**: Spatial distribution of the earthquakes generated in the synthetic catalog over the same spatio-temporal domain; grey circles and green dots show the spatial distribution of the background earthquakes and aftershocks respectively.

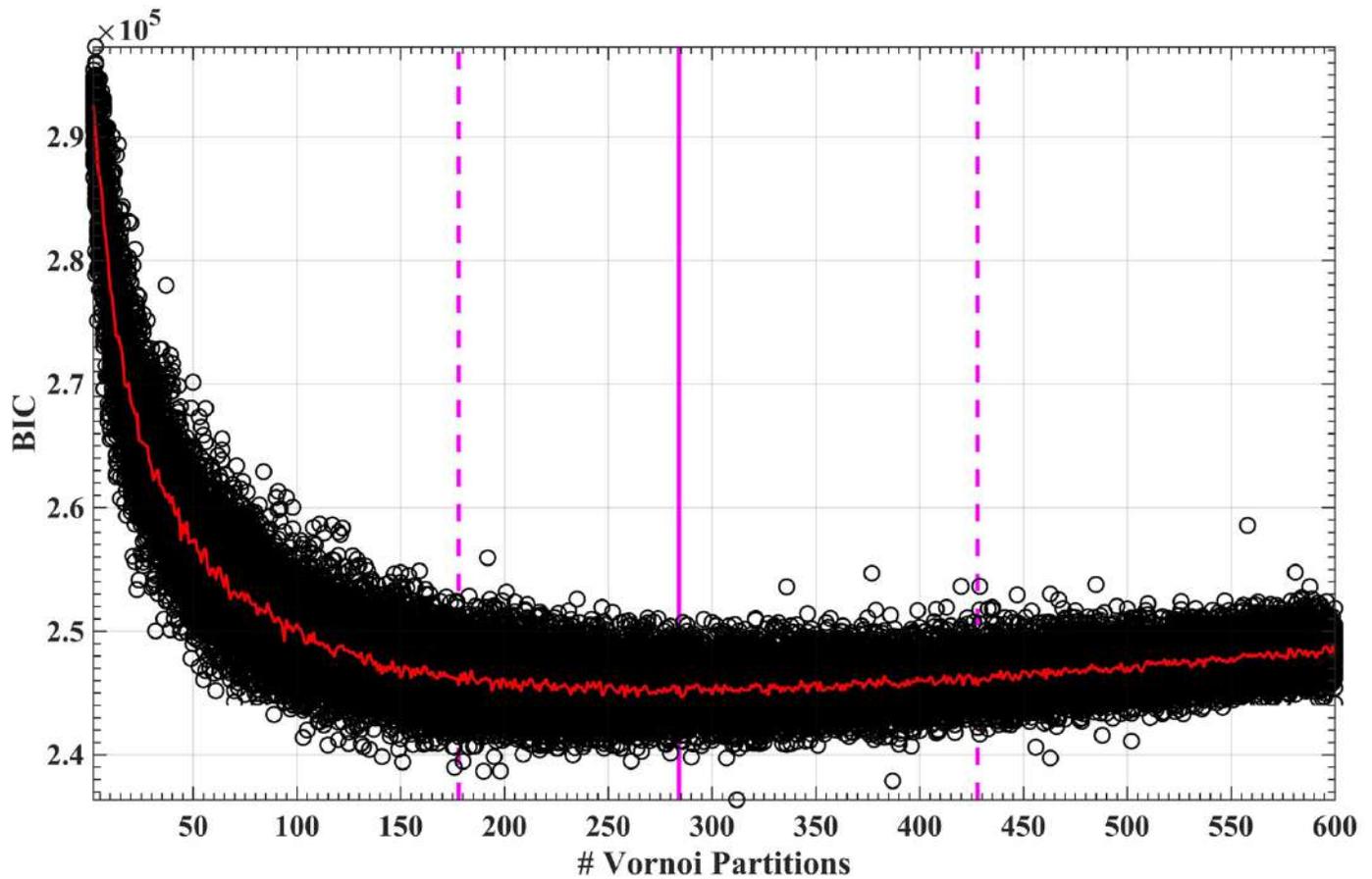

**Figure S4**: The BIC corresponding to 96,000 solutions as a function of the number of Voronoi cells used is shown using black circles; the median BIC corresponding to each Voronoi complexity level is shown using a solid red line; solid magenta corresponds to the minima in the median BIC curve (indicated using red line) and indicates the optimal complexity level; dashed magenta line indicates the optimal complexity range in which the median BIC for a given complexity level is not significantly different from the minimum median BIC

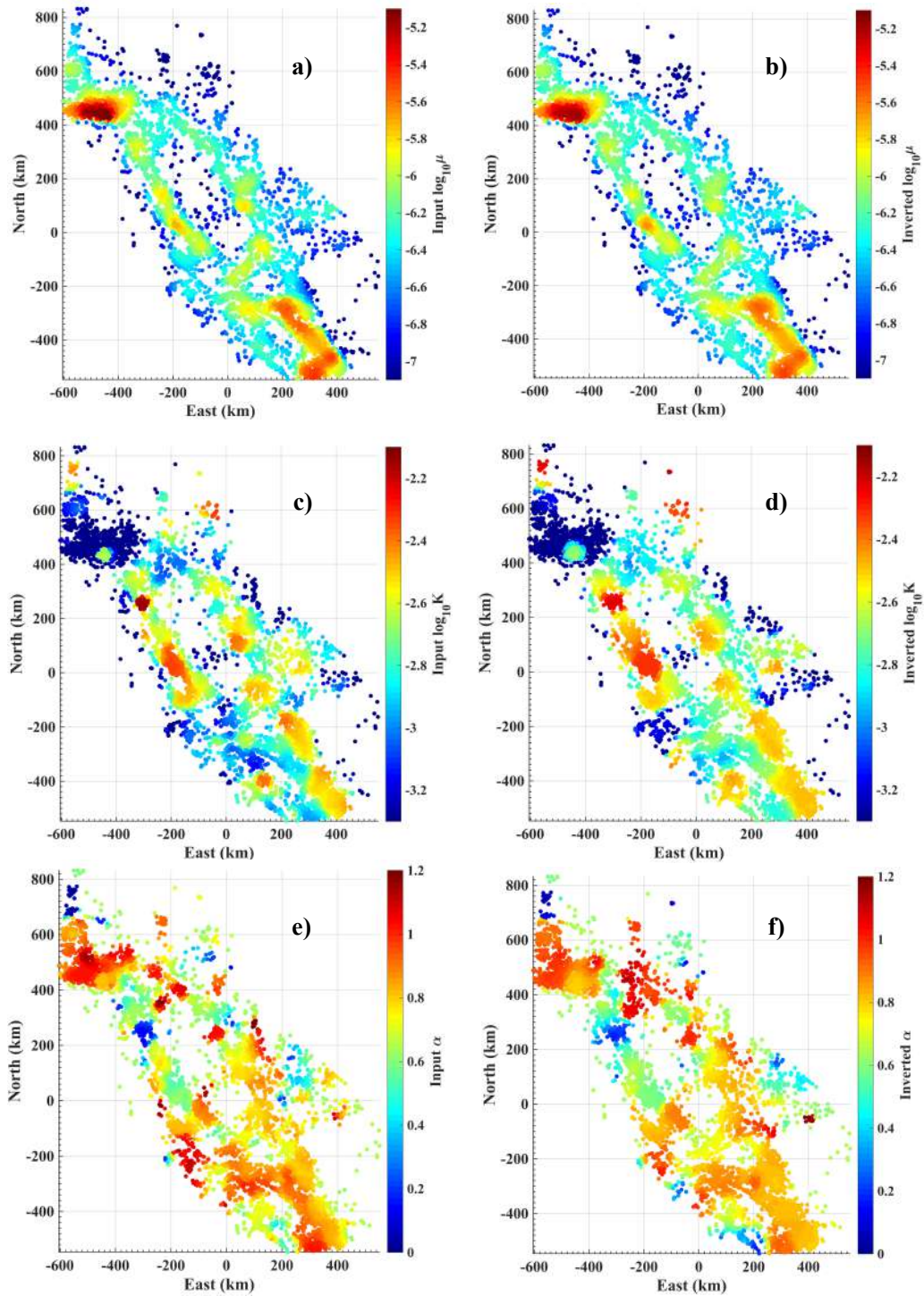

**Figure S5**: (a-f) Spatial variation of the (a) Input background seismicity rate ($\mu$, # earthquakes/$km^2/day$) (b) Inverted $\mu$ (c) Input pre-factor of the aftershock productivity (**K**) (d) Inverted **K** (e) Input exponent of the aftershock productivity ($\alpha = \frac{a}{log(10)}$) (f)

Inverted $\boldsymbol{\alpha}$; circles show the locations of the earthquakes in the synthetic catalog; colors corresponding to each earthquake in the right panels represent the ensemble estimate of $\boldsymbol{\mu}$, $\boldsymbol{K}$ and $\boldsymbol{\alpha}$ at the location of the synthetic earthquakes, computed using the solutions with the complexity range indicated by dashed magenta lines in Figure S5.

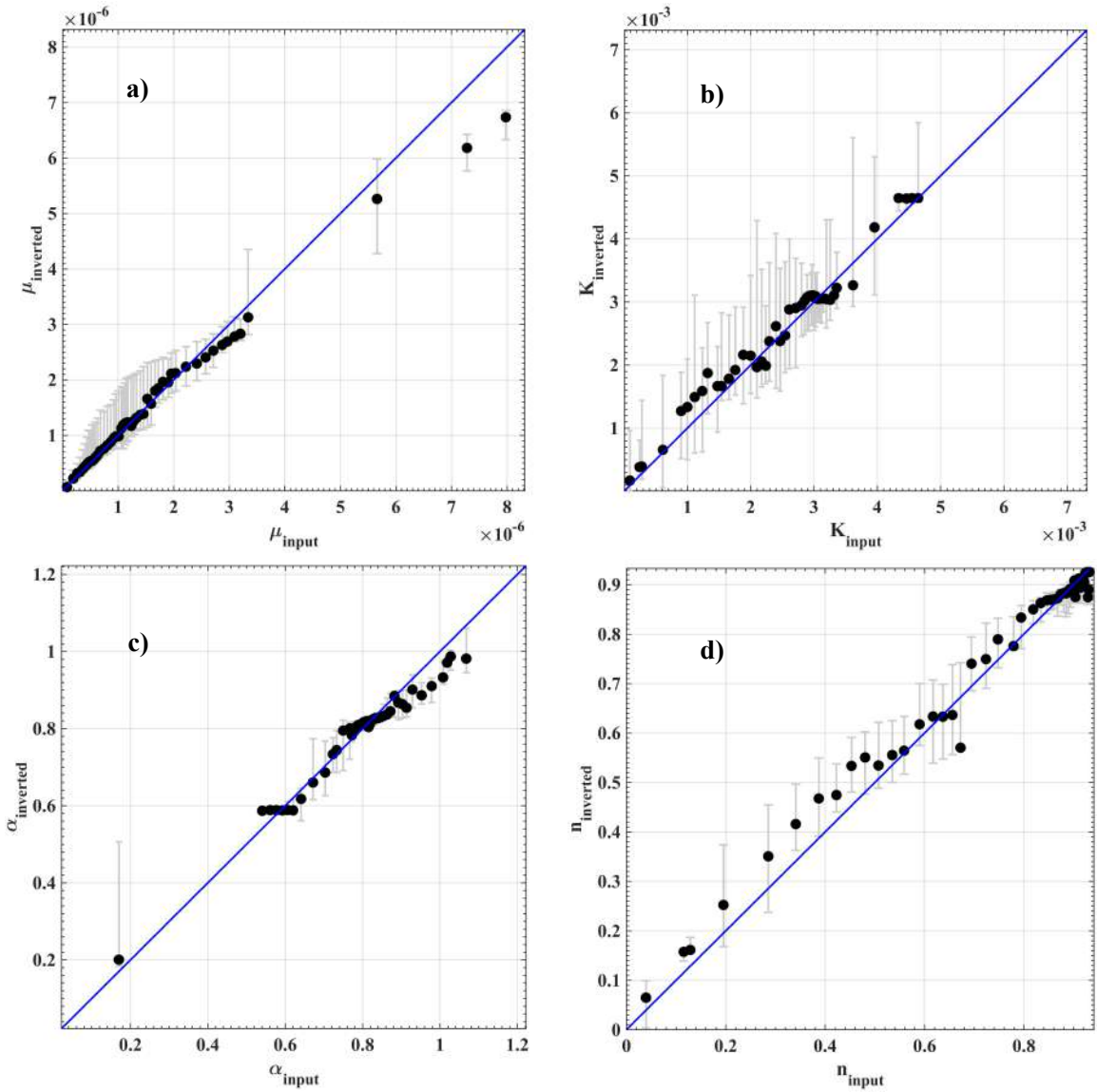

**Figure S6**: Correlation between (a) Inverted and input background seismicity rate ($\mu_{inverted}$ versus $\mu_{input}$) (b) Inverted and input prefactor of aftershock productivity ($K_{inverted}$ versus $K_{input}$) (c) Inverted and input exponent of aftershock productivity ($\alpha_{inverted} = \frac{a_{inverted}}{\log(10)}$ versus $\alpha_{input} = \frac{a_{input}}{\log(10)}$) (b) Inverted and input branching ratio ($n_{inverted}$ versus $n_{input}$); black circles show the median value and grey bars show the 95% confidence interval of Y conditioned on the median value of X; the Y=X line is shown using a blue solid line.

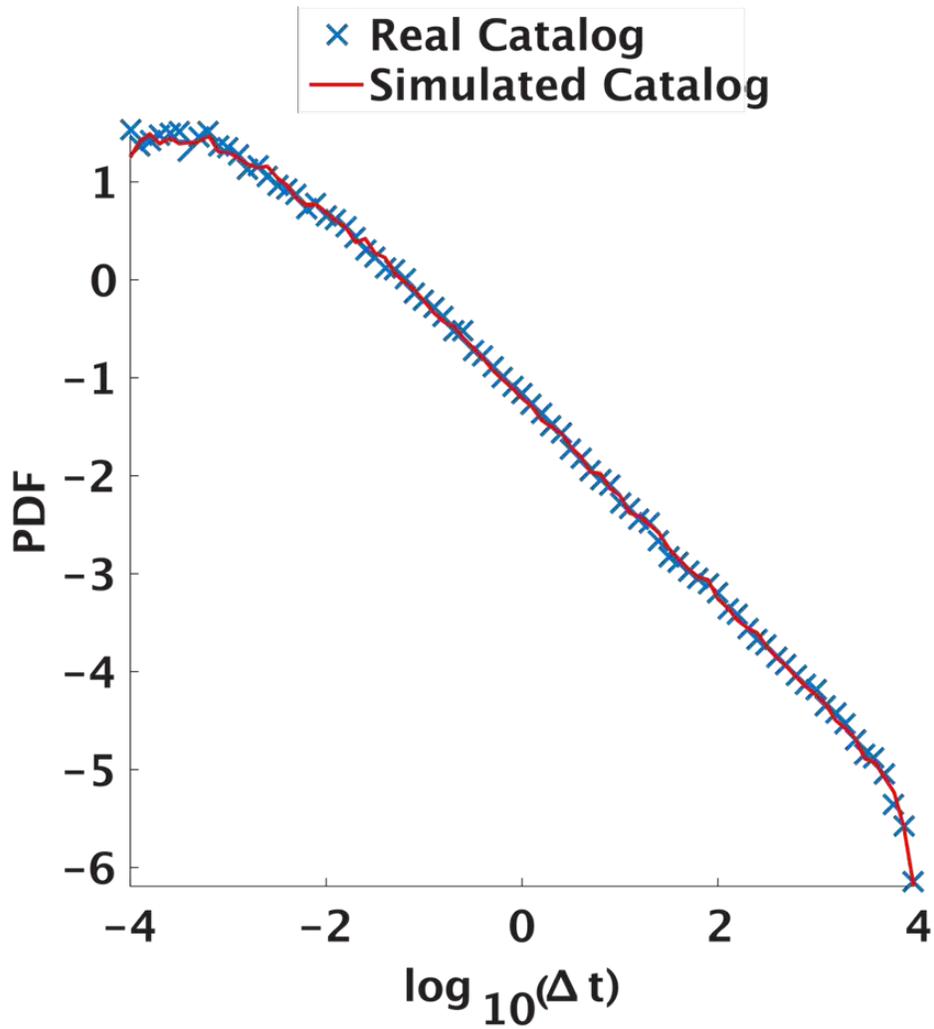

**Figure S7**: Blue crosses show the empirical PDF of waiting times between mainshocks and direct aftershocks in the real catalog; the red solid line shows the empirical PDF of waiting times between mainshocks and direct aftershocks in the synthetic catalog generated using an Omori kernel with exponent p=1.0051 and $c = 10^{-2.59}$ days.

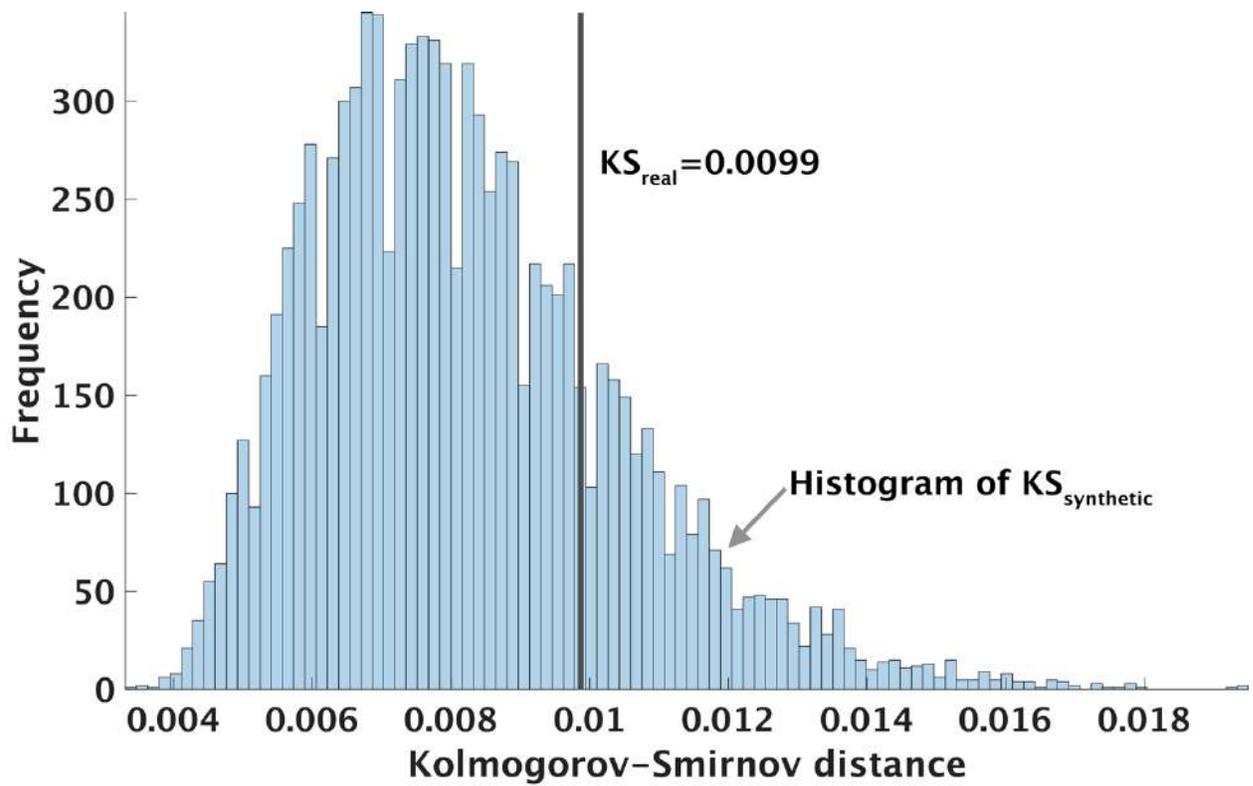

**Figure S8**: Histogram of $KS_{synthetic}$ is shown using blue columns; solid grey line shows the value of $KS_{real}$.

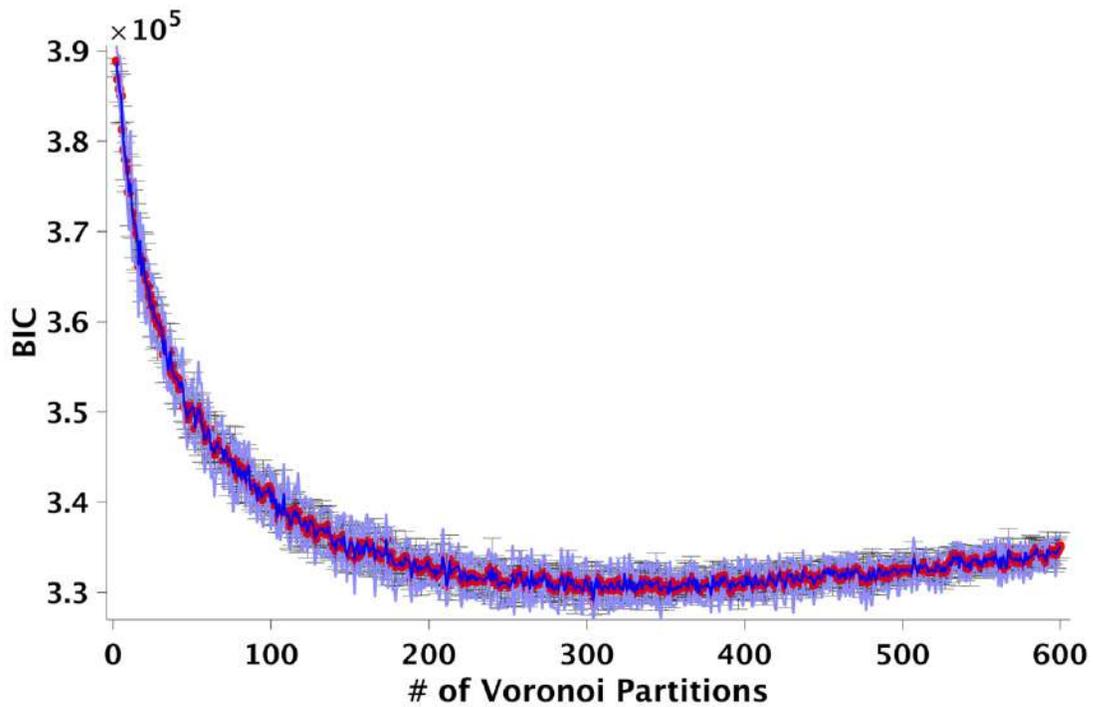

**Figure S9**: Red circles and black error bars show the median and 95% confidence interval corresponding to the modified ETAS model (ETAS-mod); solid blue lines and blue shaded region show the median and 95% confidence interval corresponding to the unmodified ETAS model (ETAS-unmod).

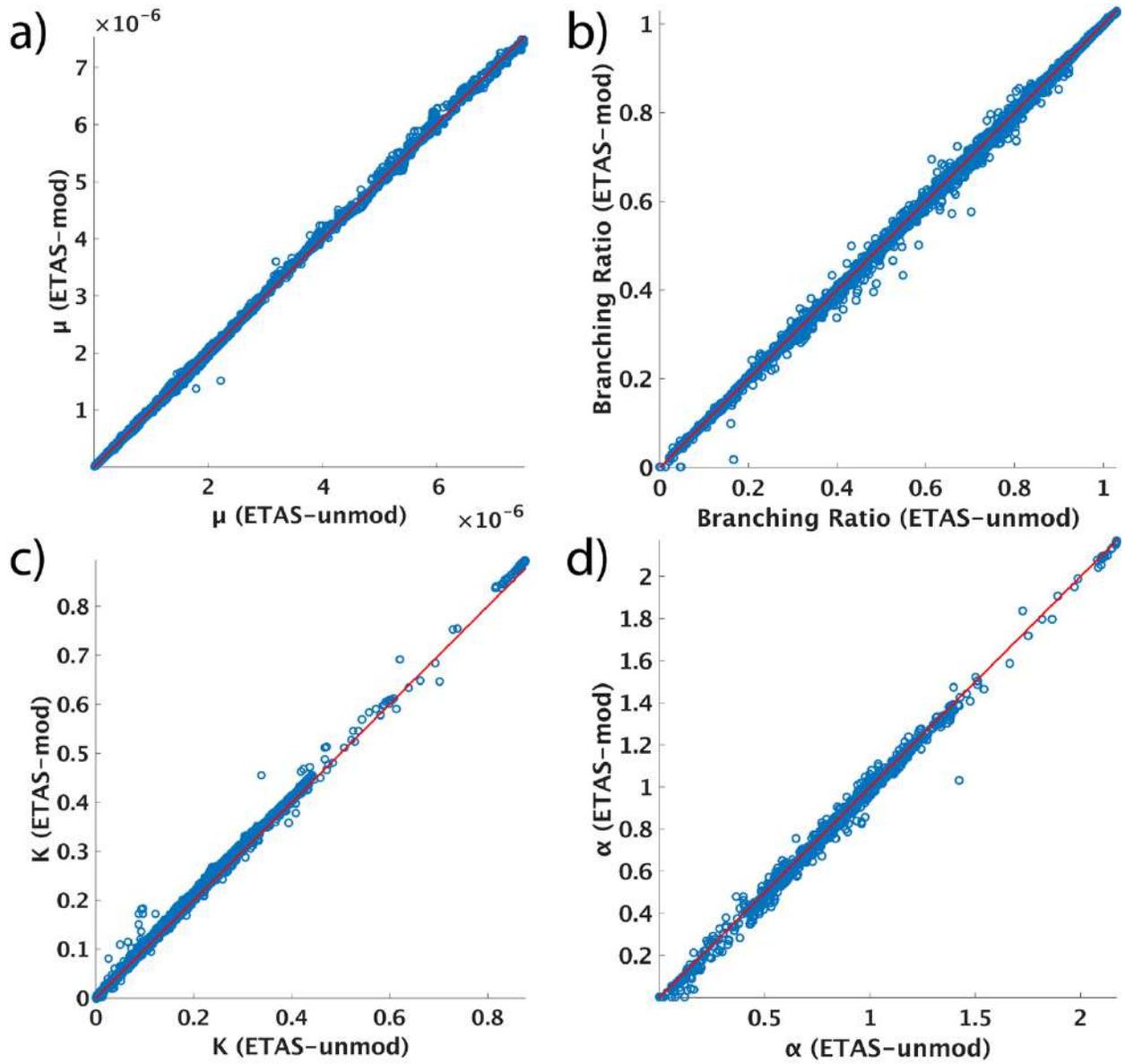

**Figure S10**: Estimate of the spatially variable parameters obtained from ETAS-mod plotted vs. ETAS-unmod models.

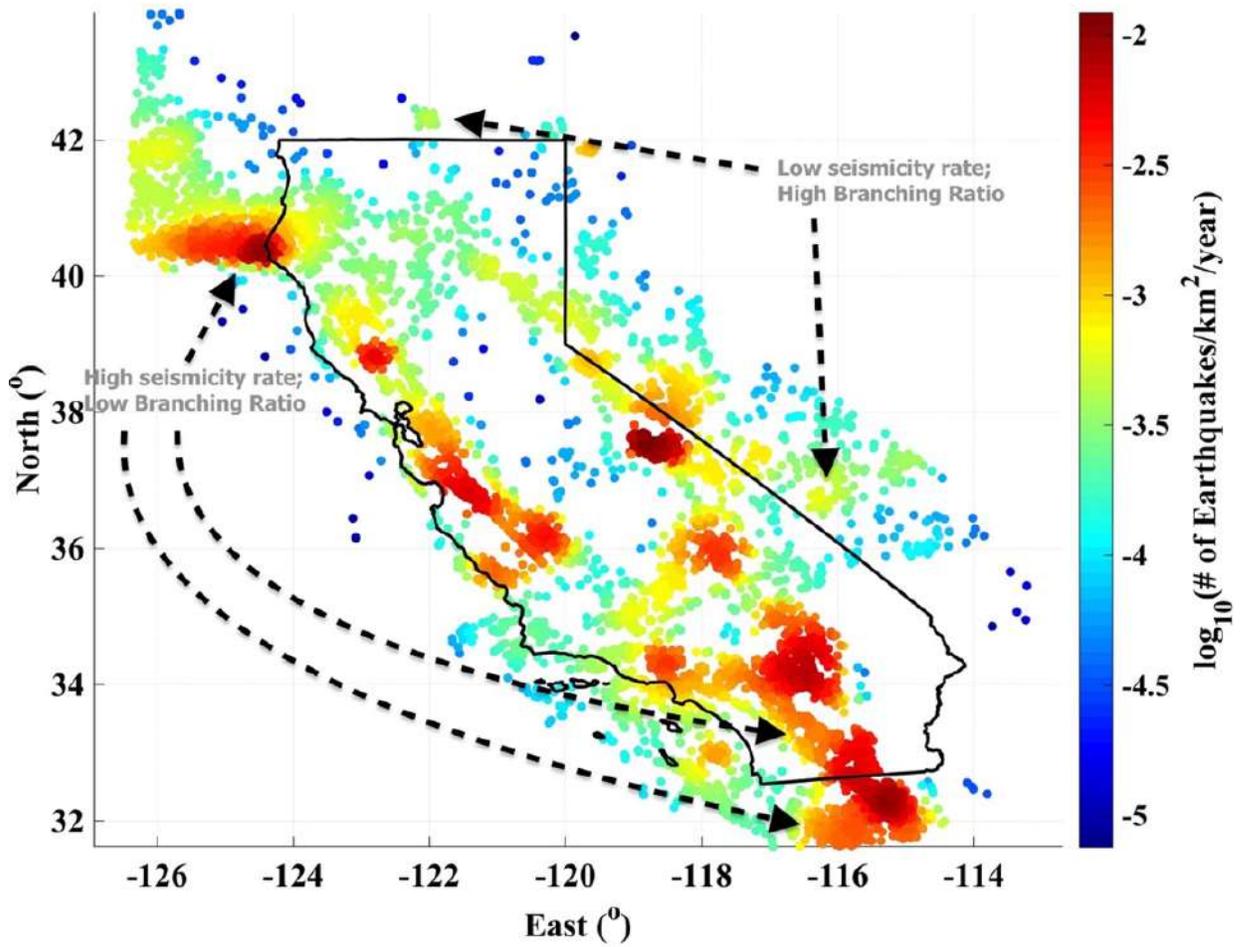

**Figure S11**: Spatial variation of the total seismicity rate ( # earthquakes/$km^2/year$); circles show the locations of the 21,448 earthquakes (M$\geq$ 3) used.

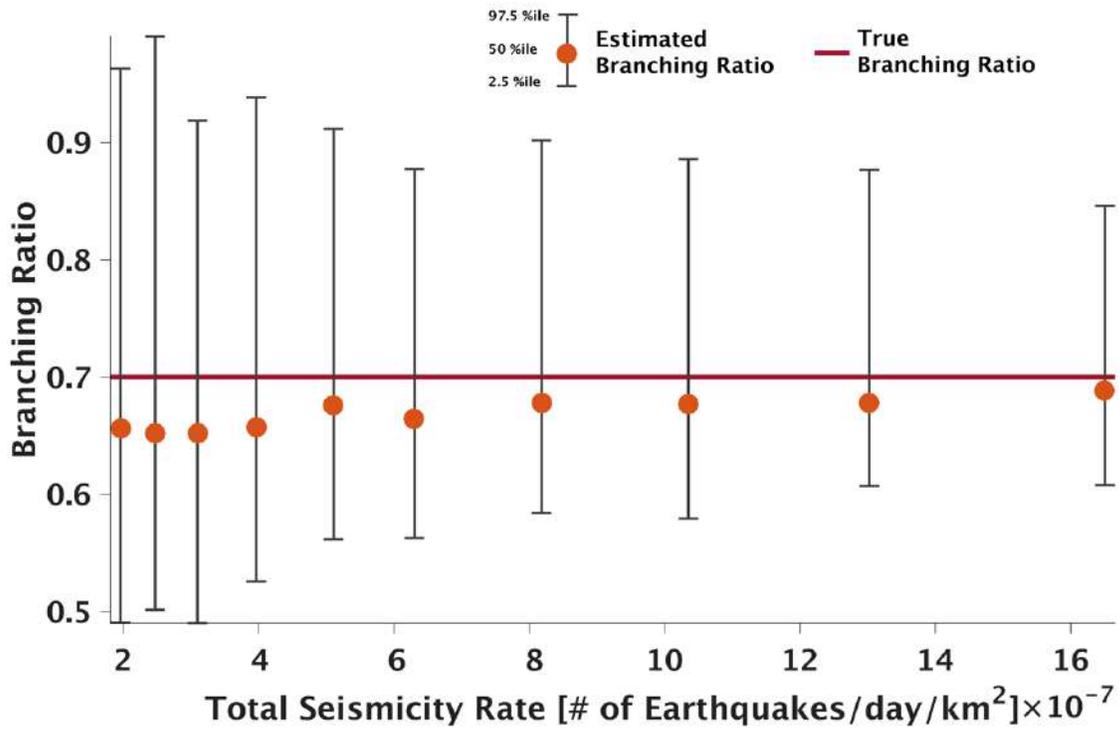

**Figure S12:** Orange circles and the error bars show the estimate of the median of the branching ratio and its 95% confidence interval for a given expected total seismicity rate; the solid red line shows the true branching ratio used to simulate the synthetic catalogs.

**Table S1:** Input and inverted values of the parameters of the spatially homogenous ETAS model

| Parameters | $log_{10}\mu$ | $log_{10}K$ | $\alpha = \dfrac{a}{log10}$ | $log_{10}c$ | $\omega$ | $log_{10}d$ | $\rho$ | $\gamma$ |
|---|---|---|---|---|---|---|---|---|
| Input value | -6.35 | -2.25 | 0.8 | -2 | 0.4 | 0.18 | 0.57 | 1.23 |
| Inverted value | -6.37 | -2.22 | 0.81 | -1.96 | 0.39 | 0.19 | 0.62 | 1.19 |

**Table S2:** Ensemble estimate of the five spatially invariant ETAS parameters; the input values of each of these parameters are shown in Figure 3a-e using solid magenta line.

| Parameters | $c$ | $\omega$ | $d$ | $\rho$ | $\gamma$ |
| --- | --- | --- | --- | --- | --- |
| **Inverted value** | $10^{-2.64}$ | 0.005 | 0.19 | 0.59 | 1.23 |

**Table S3**: Estimates of the spatially invariant parameters obtained from ETAS-unmod and ETAS-mod models.

| Models | $c$ | $\omega$ | $\eta$ | $d$ | $\rho$ | $\gamma$ |
|---|---|---|---|---|---|---|
| ETAS-unmod | $10^{-2.59}$ | 0.0051 | NA | $10^{-0.74}$ | 0.5603 | 1.2684 |
| ETAS-mod | $10^{-2.48}$ | 0.0073 | -0.1914 | $10^{-0.74}$ | 0.5591 | 1.2692 |